\documentclass[journal]{IEEEtran}
\usepackage{cite}
\usepackage{amsmath,amssymb,amsfonts}
\usepackage{algorithmic}
\usepackage{graphicx,color}
\usepackage{textcomp}
\usepackage[utf8]{inputenc}
\usepackage[table,dvipsnames]{xcolor}
\usepackage{subcaption}
\usepackage{gensymb}
\usepackage{float}
\usepackage{array}
\usepackage{hyperref}
\usepackage{eso-pic, rotating}
\usepackage{multirow}
\usepackage{multicol}
\usepackage{tabularx, makecell}
\usepackage{tabularray}
\usepackage{ctable}
\usepackage{booktabs}
\usepackage{svg}
\usepackage{diagbox}
\usepackage{glossaries}
\usepackage[flushleft]{threeparttable}

\newacronym{rat}{RAT}{Radio Access Technology}
\newacronym{rf}{RF}{Radio Frequency}
\newacronym{sei}{SEI}{Specific Emitter Identification}
\newacronym{oran}{O-RAN}{Open Radio Access Network}
\newacronym{dst}{DST}{Digital Spectrum Twin}
\newacronym{snr}{SNR}{signal-to-noise ratio}
\newacronym{ssl}{SSL}{Self-Supervised Learning}
\newacronym{cnn}{CNN}{Convolutional Neural Network}
\newacronym{sota}{SotA}{state-of-the-art}
\newacronym{ae}{AE}{Autoencoder}
\newacronym{dc}{DC}{Deep Clustering}
\newacronym{fft}{FFT}{Fast Fourier Transform}
\newacronym{ppt}{ppt}{percentage points}
\newacronym{gflops}{GFLOPS}{one billion floating-point operations per second}
\newacronym{flops}{FLOPS}{floating-point operations per second}
\newacronym{ssdc}{SSDC}{Self-supervised deep clustering}
\newacronym{lstm}{LSTM}{Long Short-Term Memory}
\newacronym{ran}{RAN}{Radio Access Network}
\newacronym{iq}{I/Q}{In-phase and Quadrature}
\newacronym{nn}{NN}{Neural Network}
\newacronym{psd}{PSD}{Power Spectral Density}
\newacronym{tsne}{t-SNE}{t-distributed Stochastic Neighbor Embedding}
\newacronym{umap}{UMAP}{Uniform Manifold Approximation and Projection}
\newacronym{dmd}{DMD}{Dynamic Mode Decomposition}
\newacronym{gasf}{GASF}{Gramian Angular Summation Field}
\newacronym{rgb}{RGB}{Red-Green-Blue}
\newacronym{nrt}{Near-RT}{Near-real-time}
\newacronym{ric}{RIC}{Radio Intelligent Controller}
\newacronym{ru}{RU}{Radio Unit}
\newacronym{dcec}{DCEC}{Deep Convolutional Embedded Clustering}
\newacronym{aeml}{AEML}{Autoencoder with Modified Loss}
\newacronym{its}{ITS}{Intelligent Transportation Systems}
\newacronym{sl}{SL}{Silhouette score}
\newacronym{db}{DB}{Davies-Bouldin score}
\newacronym{ch}{CH}{Calinski-Harabasz score}
\newacronym{nmi}{NMI}{Normalized Mutual Information}
\newacronym{ari}{ARI}{Adjusted Rand Index}
\newacronym{chs}{CHS}{Clustering homogeneity score}
\newacronym{ccs}{CCS}{Clustering completeness score}
\newacronym{pca}{PCA}{Principal Component Analysis}
\newacronym{fc}{FC}{Fully Connected}
\newacronym{tcd}{TCD-L}{Technology Classification Dataset - Labeled}
\newacronym{its_l}{ITS-L}{Intelligent Transportation Systems - Labeled}
\newacronym{unb}{UNB-U}{Ultra Narrow Band - Unlabeled}
\newacronym{isac}{ISAC}{Integrated Sensing and Communication}
\newacronym{lbt}{LBT}{Listen-Before-Talk}
\newacronym{ml}{ML}{Machine Learning}

\makeglossaries

\def\BibTeX{{\rm B\kern-.05em{\sc i\kern-.025em b}\kern-.08em
    T\kern-.1667em\lower.7ex\hbox{E}\kern-.125emX}}
\AtBeginDocument{\definecolor{ojcolor}{cmyk}{0.93,0.59,0.15,0.02}}

\begin{document}

\title{Spectrum Sensing with Deep Clustering: Label-Free Radio Access Technology Recognition\\
}


\author{
\IEEEauthorblockN{Ljupcho Milosheski\IEEEauthorrefmark{1}\IEEEauthorrefmark{2}, Mihael Mohorčič\IEEEauthorrefmark{1}\IEEEauthorrefmark{2}, Carolina Fortuna\IEEEauthorrefmark{1}}\\
\IEEEauthorblockA{
\IEEEauthorrefmark{1}Department of Communication Technologies, Jozef Stefan Institute, Jamova 39, 1000, Ljubljana, Slovenia \\
\IEEEauthorrefmark{2}Jozef Stefan International Postgraduate School, Jamova 39, 1000, Ljubljana, Slovenia \\
Email: \{ljupcho.milosheski, miha.mohorcic, carolina.fortuna\}@ijs.si
}
}

\maketitle

\begin{abstract}

The growth of the number of connected
devices and network densification is driving an
increasing demand for radio network resources, particularly
Radio Frequency (RF) spectrum.  Given the dynamic and complex nature of contemporary
wireless environments, characterized by a wide variety of
devices and multiple RATs, spectrum sensing is envisioned
to become a building component of future 6G, including as a component within O-RAN or digital twins. 
However, the current SotA research for RAT classification predominantly revolves around supervised Convolutional Neural
Network (CNN)-based approach that require extensive labeled dataset. Due to this, it is unclear how existing models behave  in environments for which training data is unavailable thus leaving open questions regarding their generalization capabilities.
In this paper, we propose a new
spectrum sensing workflow in which the model training
does not require any prior knowledge of the RATs
transmitting in that area (i.e. no labelled data) and the
class assignment can be easily done through manual
mapping. Furthermore, we adapt a SSL deep clustering architecture capable of autonomously extracting
spectrum features from raw 1D Fast Fourier Transform
(FFT) data. We evaluate the proposed architecture  on three real-world datasets from three European
cities, in the 868 MHz, 2.4 GHz and 5.9 GHz bands
containing over 10 RATs and show that the developed 
model achieves superior performance  by up to 35 percentage points with  22\% fewer trainable parameters
and 50\% less floating-point operations per
second (FLOPS) compared to an SotA AE-based
reference architecture.

\end{abstract}


\maketitle

\section{INTRODUCTION} \label{sec:intro}
The exponential growth of the number of connected devices \cite{Ericsson2023Mobility} and network densification is driving an increasing demand for radio network resources, particularly \acrfull{rf} spectrum. The allocation of new operational frequency bands, such as the 6.425-7.125 GHz range designated by the ITU for licensed mobile communications \cite{alsaedi2023spectrum}, only partially alleviates the growing demand for \acrshort{rf} spectrum resources. Thus, it is crucial to enhance the utilization of the occupied bands through innovative spectrum-sharing strategies that go beyond the existing licensed \cite{saha2024dynamic} and license-exempt access schemes \cite{parvini2023spectrum}, meeting the complexity of the fast-evolving wireless networks. Such approaches will increasingly rely on accurate monitoring and understanding of the \acrshort{rf} spectrum environment, achieved through spectrum sensing techniques. The process of spectrum sensing involves analyzing radio data for a variety of tasks \cite{wong2021rfml}, each playing an important role in the broader context of radio awareness, impacting the efficiency and effectiveness of wireless radio networks. Such tasks and corresponding purposes include \acrfull{rat} classification \cite{ke2021real} for optimizing resources allocation and spectrum utilization, modulation classification \cite{rajendran2018deep} for improving data throughput and reducing error rates, anomaly detection \cite{poornima2020anomaly} enhancing network security and resilience, interference recognition \cite{ghanney2020radio} for identification of sources of interference, and \acrfull{sei} \cite{robinson2020dilated} ensuring secure and efficient spectrum sharing by preventing unauthorized access.

Given the dynamic and complex nature of contemporary wireless environments, characterized by a wide variety of devices and multiple \acrshort{rat}s, spectrum sensing is envisioned to become a building component of future 6G networks \cite{demirhan2023integrated}. As such, it will support the progress of \acrfull{isac}-related systems and technologies \cite{ouyang2023integrated}, \cite{wei2023integrated}, facilitating their development and integration into the next generation of wireless communication systems.
Consequently, spectrum sensing is anticipated to be a key component in the evolution of emerging wireless communication frameworks such as \acrfull{oran} and \acrfull{dst} \cite{tadik2023digital}.

In the emerging \acrshort{oran} in which a large proportion of the radio functions are realized in software \cite{abdalla2022toward}, \cite{colleti2018}, \acrshort{rat} monitoring plays an important role in optimizing radio and network parameters and spectrum utilization, largely through the employment of AI algorithms encapsulated as xApps and rApps \cite{bonati2020open}. By continuously assessing the usage and availability of different \acrshort{rat}s, \acrshort{oran} can utilize intelligent applications to dynamically fine tune network settings and allocate spectrum resources or manage beam forming \cite{hoffmann2023beam} in real time. Thus, it can enhance the overall efficiency and performance of the network and accommodate growing traffic demands, while better preserving privacy and being less data intensive \cite{d2022orchestran} compared to SEI. 

\acrshort{dst}s are digital representations of the real-world characteristics of the \acrshort{rf} spectrum in a region based on historical data and measurement updates \cite{durgin2022digital}. They are envisioned as tools to inform, forecast, and enhance spectrum utilization and interference management in wireless networks in a given geographical area. Spectrum sensing is a building component of \acrshort{dst}, as outlined in \cite{tadik2023digital}. It provides the measurement input necessary to refine the \acrshort{dst}, which is initially developed based on empirical, theoretical, and/or ray-tracing models. Incorporating \acrshort{rat} monitoring into \acrshort{dst} could further enhance its capabilities by adding detailed data on the various transmission technologies that are present in the area, such as \acrshort{rat}-specific activity patterns for a given time interval within the operational landscape. This integration could significantly improve the \acrshort{dst}'s awareness, accuracy and functionality, facilitating more effective spectrum management and optimization strategies.


The current \acrshort{sota} research for \acrshort{rat} classification predominantly revolves around supervised \acrfull{cnn}-based approaches \cite{zhang2021signal, scalingi2023framework}, \cite{han2022proceed}, \cite{girmay2023technology}, which have shown promising results, particularly in diverse \acrfull{snr} settings and various environmental conditions \cite{fontaine2019towards}. However, these methods have a number of shortcomings as follows. First, they heavily depend on extensive labeled datasets acquired in controlled settings where the transmission parameters need to be recorded in addition to the received signals requiring a more complex set-up for collection compared to unlabelled data. Second,  it is unclear how they transfer to other environments characterized by (i) additional RATs compared to the original training data and (ii) different physical obstructions and noise variations. Third, the majority of the models have been developed and evaluated on a single dataset leaving open questions regarding their \textit{generalization} capabilities. 

\acrfull{ssl} approaches,  extensively explored in other application domains \cite{ren2022deep}, are able to address the first two challenges mentioned above, because they do not require labels for training. As long as a spectrum sensor collecting data is available, that data can be directly used to train such models without requiring recorded transmission parameters. By analogy, this approach should be able to adapt the model in regular retraining periods to new types of RATs that may appear in the area. However, to be able to realize the classification functionality similar to their supervised counterparts, a manual cluster-to-label assignment step is required. Nevertheless, this process is rather efficient as the learnt clusters, together with additional insights, can be presented to a human decision maker for label assignment. The decision maker is also presented with a new cluster including a new technology that emerged from the learning process and can immediately identify it. The adoption of \acrfull{ssl}  in wireless communications research remains limited. Existing \acrfull{sota} works primarily rely on variations of \acrfull{ae} architectures \cite{zhang2022cnn, han2022proceed}. These methods emphasize instance learning, where the \acrshort{cnn} part is trained solely or partially on reconstruction loss, focusing on distinguishing individual samples.

Recognizing the need for a more adaptable solution, particularly in scenarios where class information is unknown, we propose\footnote{\url{https://github.com/sensorlab/spire}} the use of a domain-adapted DeepCluster \cite{caron2018deep} architecture. This approach diverges from the existing methods by basing the learning process of the \acrshort{cnn} part on features shared among groups of samples rather than individual instances. This distinction could make the proposed architecture more general and effective in realistic environments, thus addressing the third challenge. 

The main contributions of our work are:
\begin{itemize}
    \item   \textbf{New spectrum sensing workflow:} We propose a new spectrum sensing workflow in which the model training does not require any prior knowledge of the RATs transmitting in that area (i.e. no labelled data) and the class assignment can be easily done through manual mapping.  
    \item \textbf{Development of an \acrshort{ssl} \acrfull{dc} Architecture}: We introduce an adaptation of \acrshort{ssl} deep clustering architecture capable of autonomously extracting spectrum features from raw 1D \acrfull{fft} data. Notably, this architecture is inherently environment-agnostic and adaptable to varying numbers of operating transmission technologies.
    \item \textbf{Generic Unsupervised Model}:  We go beyond the \acrshort{sota} in model development for \acrshort{rat} feature learning and subsequent classification by developing a more \textit{generic} model on three real-world datasets from three European cities, in the 868 MHz, 2.4 GHz and 5.9 GHz bands containing over 10  \acrshort{rat}s.
    \item \textbf{Efficiency and Performance Optimization}: The developed model boasts $22$\% fewer trainable parameters and requires $50$\% less \acrfull{flops} compared to an identical \acrshort{ae}-based reference architecture. Besides the efficiency gains, our model achieves superior performance in label-based clustering evaluations by up to $35$ \acrfull{ppt} on \acrshort{tcd} dataset (data of each technology acquired in six different sites), up to $4$ \acrshort{ppt} on \acrshort{its_l} dataset (samples of each technology acquired at single site) and better separation of the feature space of the unlabeled \acrshort{unb} dataset (continuously sensed data in uncontrolled environment).
\end{itemize}

The structure of this paper is as follows: Section~\ref{sec:related} reviews related work in the field. Section~\ref{sec:overview} describes the reference scenario and related assumption.
The problem formulation is detailed in Section~\ref{sec:problem}. The proposed and baseline architectures are presented in Section~\ref{sec:architecture}. Section~\ref{sec:methodology} outlines the evaluation methodology, while Section~\ref{sec:results} discusses the outcomes of this evaluation. The paper concludes with Section~\ref{sec:conclusion}, summarizing the key findings and contributions.

\section{RELATED WORK}\label{sec:related}

Considering the disadvantages of the supervised learning approaches for spectrum sensing-related tasks, such as the necessity of labels and expert intervention in the setup of models in specific environments, research work towards employing unsupervised models has attempted to advance alternative approaches that do not necessitate labels.

The performance of pure unsupervised clustering models applied to spectrum data is studied in \cite{cerar2023learning}. Authors evaluate classical unsupervised (clustering) approaches on data transformed by \acrfull{tsne} and \acrfull{umap}. Three clustering algorithms are subjected to comparison: K-means, Agglomerative Hierarchical Clustering, and Hierarchical Density-Based Spatial Clustering. Although such an approach benefits from being general and simple for high-level spectrum analysis, feature extraction is based on \acrshort{tsne} and \acrshort{umap}, which is disadvantageous compared to \acrshort{cnn}-based, deep feature learning.

A study of another lightweight and general approach for \acrshort{rat} classification is proposed in \cite{elsebaay2023wireless} as an alternative to models based on \acrshort{cnn}. It introduces \acrfull{dmd} as a feature extraction technique capable of distinguishing \acrshort{rat}s based on their bandwidth. The classification process involves a straightforward threshold method, demonstrating superior performance compared to a baseline \acrshort{cnn}-based solution that operates on \acrfull{gasf} images derived from time series data. Although the solution benefits from its simplicity, it faces the problem of manual adjustment of thresholds.

\acrshort{ae}s are neural networks trained to encode input data into a compressed representation and then decode that representation back to an approximation of the original input. They are primarily used for dimensionality reduction or feature learning. However, in their original form, where learning is based on reconstruction loss only, they do not provide clustering-friendly embedded space. This led to research on modifications of this architecture by adding different loss functions, as discussed in \cite{meng2023unsupervised}.
\acrshort{ae} with an additional clustering loss (\cite{ma2019learning}, \cite{zhou2023deep}, \cite{guo2017deep}) in the embedded space is an \acrshort{ssl} model designed for both dimensionality reduction and clustering tasks. Such a model combines the feature-learning capabilities of \acrshort{ae}s with the grouping intuition of clustering algorithms by optimizing for a representation that serves both purposes. In general, the clustering loss in the embedded space could be calculated based on distances between samples or based on their distribution.

In \cite{zhou2023deep}, a deep learning, \acrshort{ae}-based solution is employed for unsupervised feature learning and clustering of embedded features on radio data for the task of modulation recognition. Alongside the \acrshort{ae}-specific reconstruction loss, the authors introduce a "Deep Clustering" loss in the embedded space. The total loss function aims to simultaneously improve reconstruction and clustering in the embedded space. The authors provide a detailed evaluation of their proposed model, considering that the number of clusters is the same as the number of different modulations, which is an optimistic setup because it requires knowledge of the exact number of modulations that exist in the data. In a real-world \acrshort{rat} monitoring scenario, it is much more realistic to assume an unknown number of classes/\acrshort{rat}s (modulations in their case).

\acrfull{dcec} is another modification of \acrshort{ae} for deep feature learning, initially dedicated to image processing proposed in \cite{guo2017deep}. In this work, the total loss function is a combination of the typical reconstruction loss and Kullback-Leibler divergence loss in the embedded space. While similar learning \acrshort{cnn} modules could be used in both approaches, the different loss calculations may optimize the convolution filters to focus on different features in the input data.

Our previous work \cite{milosheski2023self} involves a high complexity self-supervised model as it operates on waterfall plots/spectrograms (2D image-like matrices) employing off-the-shelf ResNet18 model, initially developed for \acrfull{rgb} images. This approach provides valuable information about the time-frequency occupancy of the spectrum with different \acrshort{rat}s that operate in the observed band. Although a substantial reduction of complexity of the feature vectors is achieved by augmenting the feature processing, the model itself keeps the same (high) complexity as its original implementation for feature learning of \acrshort{rgb} images. While this could be advantageous, considering it is off-the-shelf, proven architecture, the high complexity could prevent application in the lower control layers of the \acrshort{oran}, for example, as xApp in the \acrfull{nrt} \acrfull{ric} due to the short control loop times.

In this manuscript, we propose a 1D-CNN-based adaptation of the DeepCluster architecture for feature learning and unsupervised classification of RATs by processing 1D FFT data instead of 2D spectrograms. While the loss of time-frequency dependency is a drawback, this issue can be mitigated through post-processing of the classified 1D swipes.

\begin{figure}[hbt!]
  \centering
  \includegraphics[width=\columnwidth]{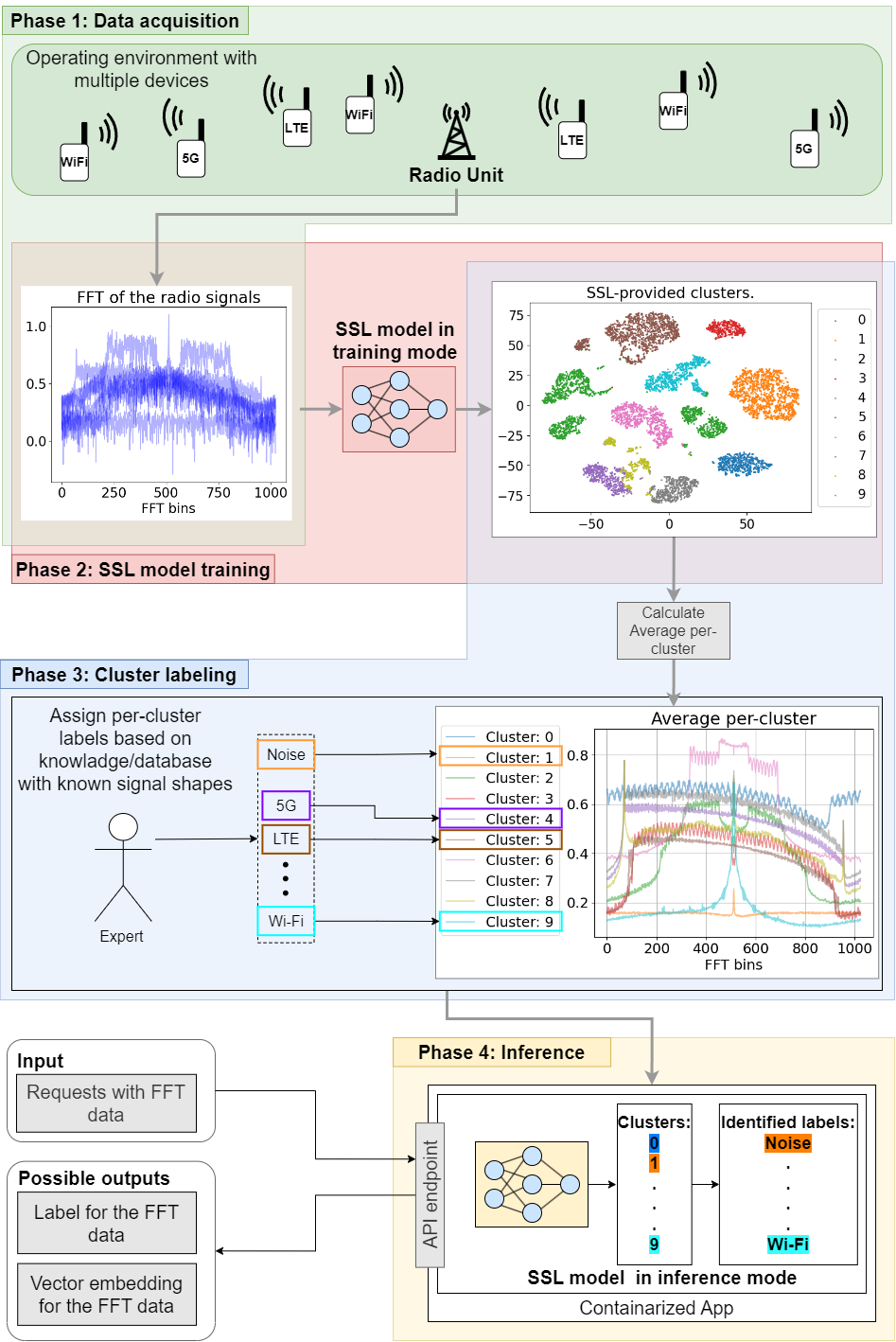}
  \caption{\acrshort{ssl} enabled spectrum sensing workflow.}
  \label{fig:SystemDescription}
\end{figure}

\section{REFERENCE SCENARIO AND PROPOSED SPECTRUM SENSING  WORKFLOW} 
\label{sec:overview}

In this paper, we assume a realistic scenario in which terminals, equipped with various RATs as depicted at the top of Figure \ref{fig:SystemDescription}, operate in a given area, possibly also in the same frequency band. For instance, a spectrum sensor may operate in ITS 5.9~GHz band where LTE, Wi-Fi, 5G NR, C-V2X PC5, and ITS-G5 technologies may co-exist \cite{girmay2023technology}. Other examples may refer to monitoring the US 6~GHz band, where Wi-Fi 6E and 5G New Radio Unlicensed (NR-U) share the unlicensed spectrum, or the LTE/5G and military radar that are sharing the 150~MHz CBRS band at 3.5~GHz \cite{luo2023spectrum}.

Furthermore, we work under the assumption that we have no prior information on what kind of RATs exist in the area under observation. Without having any prior knowledge on the transmitters in the sensed spectrum, we aim to identify them and recognize them in the future. In order to reach our aim, we propose a four-step workflow as depicted in Figure \ref{fig:SystemDescription} and elaborated below.

\subsection{Phase 1: Data aquisition}
\label{sec:phase1}
The Phase 1, marked with the green area in Figure \ref{fig:SystemDescription}, represents the data acquisition process. In this phase, the Radio Unit collects data from the operating environment in the form of the \acrshort{fft} transformation of the captured radio signals.

\subsection{Phase 2: SSL model training}
\label{sec:phase2}
The model training, marked with red in Figure \ref{fig:SystemDescription}, uses the data collected in Phase 1 and learns to group similar FFT shapes together. As a result of the grouping, clusters of similar shapes emerge with each cluster ideally containing samples from a single \acrshort{rat}. As the output of this phase, we obtain the trained model for feature extraction from \acrshort{fft} data and formed clusters.

\subsection{Phase 3: Cluster labelling}
\label{sec:phase3}
In this phase, a human expert is able to recognize the existinf RATs by looking at the clusters and manually assign a label, i.e. specific RAT such as LTE, WiFi, etc., to each cluster. In this step, hours, days or weeks of spectrum sensing are summarized for the expert end-user to quickly glance at and understand what is happening in the monitored environment. The summary can be presented as an average per cluster sweep such as depicted in Figure \ref{fig:SystemDescription} or as lower dimensional projections such as T-SNE.

\subsection{Phase 4: Inference}
\label{sec:phase4}
The model trained on real-world samples in Phase 2 together with the per cluster labels provided in Phase 3 can be packaged as a classification service where the human-assigned labels represent the class of the sample assigned to a cluster. Therefore, the classifier can be encapsulated as a containerized application with an API endpoint, and deployed as a containerized application in a production level system where incoming sweeps would get labels assigned. Furthermore, the trained model could also be exposed as an embedding model, where a compact and low-dimensional representation of the input data is provided to the requester as illustrated in Phase 4 of Figure \ref{fig:SystemDescription}.

\section{PROBLEM STATEMENT }
\label{sec:problem}

In this paper, we formulate the spectrum sensing problem as a clustering problem in machine learning that partitions the set of electromagnetic energy measurements collected during individual sweeps into groups (i.e. clusters) that contain sweeps with similar shapes (i.e. energy level envelope). These are later identified and labeled by an expert. This corresponds to Phases 1-3 described in Section~\ref{sec:overview} and depicted in Figure \ref{fig:SystemDescription}. We represent a set of electromagnetic energy measurements corresponding to a sweep as $\mathbf{X} = \{\mathbf{x}_1, \mathbf{x}_2, \ldots, \mathbf{x}_n\}$ and the set of identified \acrshort{rat}s as $\mathbf{L} = \{L_1, L_2, \ldots, L_k\}$. To account for realistic environments where the number of types of transmissions is not known apriori, the number of clusters $k$ is assumed to be unknown. This assumption uniquely distinguishes this work from the SotA where the number of classes or clusters are assumed apriori. The objective is to learn a composite function $ \Phi: X \rightarrow Y$ that maps raw data points to their identified classes in the clustered space.

\begin{equation}
    \label{mathref:classification}
    L=\Phi (X)
\end{equation}

This function can be realized by decomposition into three sub-functions, the embedding function $\phi$, the clustering function $\psi$, and the mapping function $\mathcal{M}$ and can be expressed as:
\begin{equation}
    L = \mathcal{M}(\psi(\phi(X)))
\end{equation}

\subsection{1. Embedding function} \label{sec:encoding_func}

Denoting the $\phi$ as the embedding function that transforms raw data $\mathbf{X}$ into an embedded space and $Z = \{z_1, z_2, \ldots, z_n\}$ as the set of embedded representations of the data points, formally it is defined as:
\begin{equation} \label{mathref:encoding-func}
    \phi : X \rightarrow \mathbb{R}^d
\end{equation}
where $\phi(x_i) = z_i$ and $d$ is the dimensionality of the embedded space.

\subsection{2. Clustering Function} \label{sec:clustering_func}

Considering $C = \{C_1, C_2, \ldots, C_K\}$ as the set of clusters and $K$ as the number of clusters where $K \le n$, the clustering function $\psi$ that assigns each embedded data point from $\mathbf{Z}$ to a cluster can be defined as:
\begin{equation} \label{mathref:clustering-func}
    \psi : \mathbb{R}^d \rightarrow \{1, 2, \ldots, K\}
\end{equation}
Where $\psi(z_i) = k$ indicates that the embedded point $z_i$ is assigned to cluster $C_k$.

\subsection{3. Mapping Function} \label{sec:mapping_func}

Considering $L = \{l_1, l_2, \ldots, l_K\}$ as the set of expert identified labels, the mapping function denoted as $\mathcal{M}$, can be defined as:
\begin{equation} \label{mapping-func}
\mathcal{M} : \{1, 2, \ldots, K\} \rightarrow L
\end{equation}
Where $\mathcal{M}(k) = l_j$ means that cluster $C_k$ is mapped to the label $l_j$.

\begin{figure*}[hbt!]
  \centering
  \includegraphics[width=\textwidth]{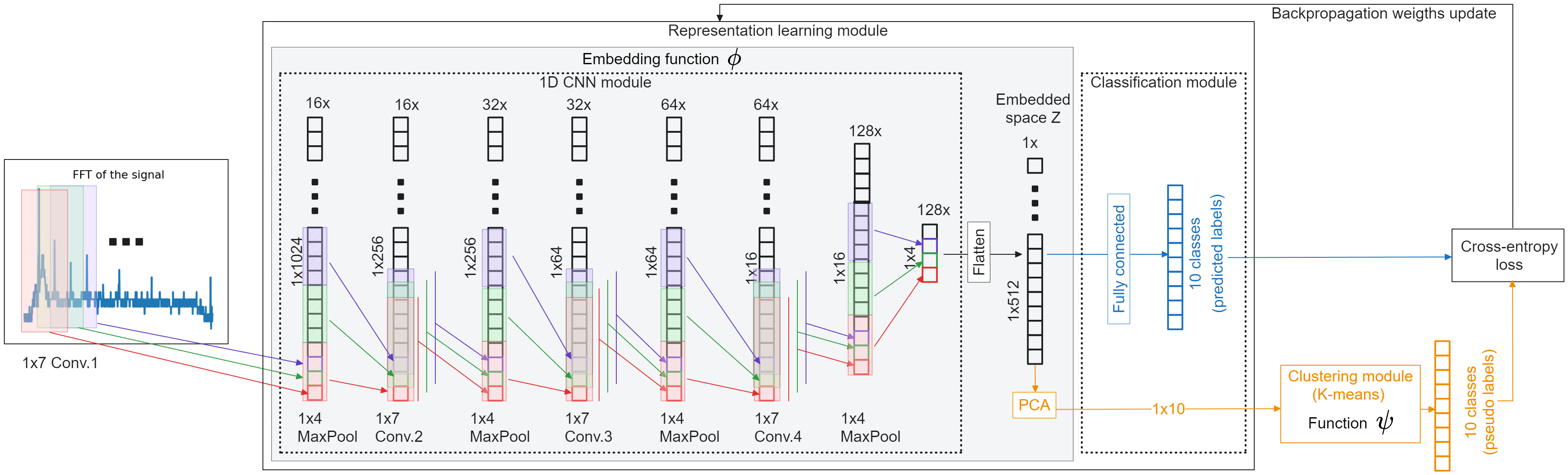}
  \caption{Proposed \acrshort{ssdc} architecture.}
  \label{fig:proposed}
\end{figure*}

\subsection{Constraints}
The constraint lies in the unknown number of clusters,
$\mathbf{K}$, for which the clustering function $\psi$ should be modeled. This is due to the uncontrolled nature of the monitored environment, where there is no prior information about the number of operating \acrshort{rat}s. This scenario is general and mirrors a real-world deployment where, regardless of the operating band, new types of signals could appear during the operation of the system, either coming from unauthorized/unknown \acrshort{rat}s in the licensed bands or from unknown \acrshort{rat}s in the unlicensed bands.

\section{PROPOSED SSL ARCHITECTURE}
\label{sec:architecture}

To solve the problem formalized in Section~\ref{sec:problem}, we propose a \acrshort{ssl} deep clustering architecture depicted in Figure~\ref{fig:proposed}. The architecture iteratively combines deep learning with clustering to learn meaningful data representations without labeled samples. The embedding function from Eq.~\eqref{mathref:encoding-func} is realized by the \acrshort{cnn} module, part of the representation learning module, while the clustering function from Eq.~\eqref{mathref:clustering-func} is realized by the clustering module.

\subsection{Description of architectural modules} \label{sec:ssdc_components}

\subsubsection{Clustering module} \label{sec:clus_module}

The clustering module groups similar data points in the embedded space. To realize the clustering module, which in turn realizes the clustering function $\psi$, we chose the well-known K-means algorithm with the Euclidean distance metric. K-means has been shown to perform very well with embedded spaces \cite{macqueen1967some}, is easy to understand, and has low computational complexity. The Euclidean metric used as a loss function for the clustering is also computationally efficient, making it appropriate for large datasets, and it is also effective for lower-dimensionality embedded spaces \cite{Aggarwal2001}. It measures the straight-line distance between data points, which allows for a clear interpretation of how data points are assigned to clusters based on their proximity in the feature space.

\subsubsection{Representation learning module} \label{sec:learning_module}

The role of the representation learning module is to learn the mapping from the input raw data to the pseudo-labels generated by the clustering process. The CNN part of the module, together with the flattening and PCA transformation, realizes the embedding function $\phi$, which transforms each input data point into a low-dimensional representation in the embedded space.

The \acrshort{cnn} module of the \acrshort{ssdc} architecture consists of four 1D convolutional layers, each followed by batch normalization, max-pooling, and \textit{ReLU} activation layers, visualized in Figure \ref{fig:proposed}. Batch normalization typically improves convergence speed \cite{ioffe2015batch}, max-pooling reduces the dimensionality of the data through the layers while focusing on the most prominent features, and ReLU is important for catching the nonlinear dependencies in the data. Regarding the number and size of \acrshort{cnn} filters, we followed the design principles discussed in \cite{o2018over} and \cite{simonyan2014very}, consisting of a stack of convolution layers followed by fully connected layers.  Each consequent convolutional layer consists of double the number of filters of the previous layer and a vector size that is four times smaller than that of its predecessor. We increase the filter size to $7$ (compared to 3 in \cite{o2018over}) so its size is large enough to neglect noise influence but small enough to be sensitive to the changes induced by the \acrshort{rat}-specific content in the \acrshort{fft} amplitudes.

Such an architecture totals $128,406$ parameters and 0.2 \textit{\acrshort{gflops}}, which is significantly lower (up to approx. 100 times, i.e. two orders of magnitude) compared to the $11.7$ million parameters and (9 times) 1.81 \textit{\acrshort{gflops}} of a model dedicated for spectrograms processing in \cite{milosheski2023self}.

\begin{figure*}[ht]
  \centering
  \includegraphics[width=\textwidth]{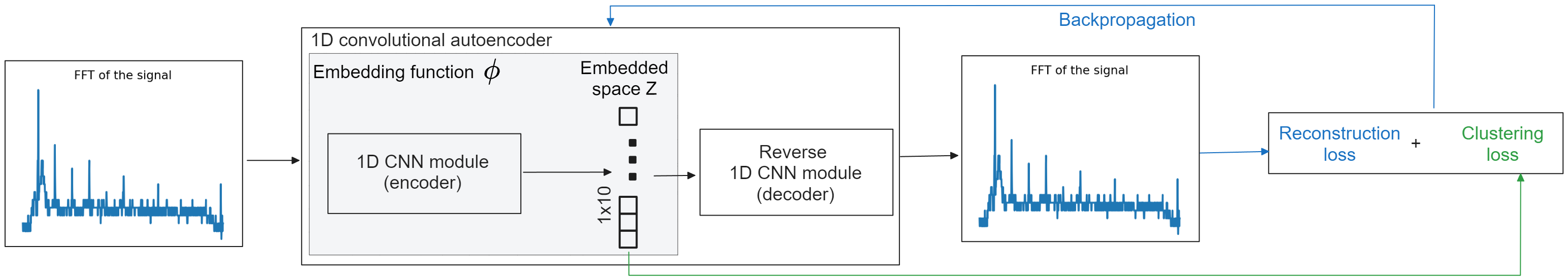}
  \caption{SotA autoencoder-based architecture \cite{zhou2023deep}.}
  \label{fig:SoA}
\end{figure*}

\subsection{Workflow of the SSDC architecture.} \label{sec:ssdc-workflow}
The \acrshort{ssdc} follows an iterative workflow consisting of two branches executed successively, mutually optimizing the clustering module and the weights of the representation learning module. The working of the clustering branch (marked with orange color in Figure \ref{fig:proposed}) consists of:

\begin{enumerate}

    \item The 1D \acrshort{cnn} module, which is part of the embedding function $\phi$, and is set to evaluation mode, having randomly initialized weights. This module extracts features from the input data, processing the raw \acrshort{fft} through the convolutional layers and providing embedded $512 \times 1$ vector representations of the samples at its output. The embedded representation is further transformed to an even lower dimensionality of $1 \times 10$ using PCA transformation \cite{bengio2013representation}. The goal is to preserve only the high-variety features of the embedded space before passing it to the clustering module. The \acrshort{cnn} module, the flattening layer, and the \acrshort{pca} realize the embedding function defined in Eq.~\eqref{mathref:encoding-func}.
    \item The feature vectors serve as an input to the clustering module, realizing the clustering function $\psi$ in our architecture. The K-means assigns pseudo-labels to the processed samples according to:
    \begin{equation} \label{mathref:cluster}
        g(\mathbf{z}_i) = \arg \min_{j} \|\mathbf{z}_i - \mathbf{\mu}_j\|,
    \end{equation}
    where \( g \) represents the cluster assignment function, \( \mathbf{z}_i \) stands for the feature vector, and \( \mathbf{\mu}_j \) is the centroid of cluster \( C_j \). The centroid \( \mathbf{\mu}_j \) is defined as:

    \begin{equation}
        \label{mathref:centroids}
        \mathbf{\mu}_j = \frac{1}{|C_j|} \sum_{\mathbf{z}_i \in C_j} \mathbf{z}_i.
    \end{equation}
    
    The K-means clustering forms clusters by minimizing the within-cluster sum of squares, which is given by:
    
    \begin{equation}
        \label{mathref:clusterloss}
        \min_{\mathbf{C}, k} \sum_{j=1}^{k} \sum_{\mathbf{z}_i \in C_j} \|\mathbf{z}_i - \mathbf{\mu}_j\|^2.
    \end{equation}
    The output of the K-means algorithm is the partitioned set of feature vectors \( \mathbf{Z} = \{\mathbf{z}_1, \mathbf{z}_2, \ldots, \mathbf{z}_n\} \) into \( k \) clusters. Ideally, each cluster \( C_j \subseteq \mathbf{Z} \) would represent a single \acrshort{rat}. The provisioning of the pseudo-labels ends the clustering module optimization as part of a single iteration.
\end{enumerate}

The representation learning module is set to the train mode in the second branch of the iteration, with the corresponding flow in Figure \ref{fig:proposed} marked with blue. Considering the pseudo-labels provided by the clustering branch described above, the weights of the representation learning module (including the \acrshort{cnn} and the \acrshort{fc} part) are trained in a classical supervised learning procedure.
\begin{enumerate}

    \item The representation learning \cite{bengio2013representation} module maps each of the input raw \acrshort{fft} samples to a single class at the output of the classifier, minimizing the difference between the pseudo-labels and the predicted labels using the Cross-entropy loss function defined as:
    
    \begin{equation} \label{mathref:cross-entropy}  
        \mathcal{L}_{\text{CE}} = -\sum_{i=1}^{n} \sum_{k=1}^{K} \mathbf{1}_{[g(\mathbf{z}_i) = k]} \log p_{i,k}.
    \end{equation}

    \item Parameters $\theta$ are updated using gradient descent and backpropagation according to
    \begin{equation} \label{mathref:weights-update}
        \theta \leftarrow \theta - \eta \nabla_{\theta} \mathcal{L}_{\text{CE}},
    \end{equation}

    where $\eta$ is the learning rate, and $\nabla_{\theta} \mathcal{L}_{\text{CE}}$ represents the gradient of the loss function with respect to the network parameters $\theta$.
    
\end{enumerate}

The two separate procedures for optimizing the clustering and the representation learning module complete one full learning iteration. 
In summary, the clustering algorithm groups input data based on their distances in the embedded domain, providing pseudo-labels. The \acrshort{cnn} and classification modules learn this distribution by enhancing the extracted features and predicting the pseudo-labels with each iteration. This iterative process between the \acrshort{cnn}-based classification and clustering modules continues for a predefined number of training epochs, starting with randomly initialized \acrshort{cnn} weights.

\subsection{REFRENCE AUTOENCODER-BASED ARCHITECTURES}
\label{sec:ae_models}

We compare our proposed \acrshort{ssdc} architecture with two variants of \acrfull{ae}-based \acrshort{ssl} architecture, depicted in Figure \ref{fig:SoA}. The \acrshort{ae}-based architecture is composed of three main modules: encoder, embedding layer, and decoder. Regarding the problem formulation in Section~\ref{sec:problem}, in this architecture, the encoder module represents the embedding function, and the clustering function works on the embedded space representations. Input for the encoder module is the raw data, which is processed by the sequential convolutional layer filters and converted into low-dimensional representation. This low-dimensional representation is flattened and transformed into a 1D vector representation in the embedding layer, which is in the middle of the architecture. The decoder block, which follows the embedding layer, has the same number of layers as the encoder and works in the opposite direction, increasing the dimensionality of the feature vector to the original size of the raw data. Providing the same dimensionality of the data at the output enables reconstruction loss in the learning process. In our work, we purposely use the same \acrshort{cnn} module that was used in the \acrshort{ssdc} architecture for building the \acrshort{ae}-based architectures. This provides equal ground for comparison that will highlight the performance differences which arise from the architecture itself, and not from the variation in the trainable parameters in the representation learning module. However, the \acrshort{ae}-based architectures, which are used as baselines, have different loss functions in the embedded space:
\begin{itemize}
    \item \acrfull{aeml} has a custom loss function in the embedded domain based on the relative distances between the samples, as proposed in \cite{zhou2023deep}.
    \item \acrfull{dcec} utilizes the Kullback-Leibler divergence loss in the embedded domain concerning the distribution of the samples instead of their relative distances. \acrshort{dcec} was designed for image processing in \cite{guo2017deep}.
\end{itemize}

\subsubsection{Workflow of the AE-based architectures}

The training workflow of the reference \acrshort{ae}-based models, \acrshort{aeml} and \acrshort{dcec}, consists of two phases, pretraining and joint training.

\paragraph{Pre-training}
The pretraining phase is the same as for the regular autoencoder. The encoder part takes high-dimensional input data and compresses it into a lower-dimensional embedded space (latent space), and the decoder part aims to reconstruct the original input data from the compressed form obtained by the encoder without considering the clustering loss. The goal is to produce a reconstruction as close as possible to the original input, thereby ensuring that the embedded space captures the essential features of the data.

During the pre-training phase, the autoencoder is trained to minimize the reconstruction error without considering the clustering loss. Thus, the weights update is performed using the reconstruction loss, which is the mean squared error (MSE) between the input \( x \) and its reconstruction \( \hat{x} \), expressed as:

\begin{equation} \label{mathref:mse-loss}
    \mathcal{L}_{\text{recon}} = \frac{1}{n} \sum_{i=1}^{n} \| x_i - \hat{x}_i \|^2
\end{equation}
where \( x_i \) is the input data point, and \( \hat{x}_i \) is the reconstructed data point.

\paragraph{Joint Training}
In this second phase, the model is trained with a combined objective function that includes both the reconstruction loss, calculated as the mean squared error between the input and its reconstruction, and the clustering loss, i.e., a loss that measures how well the clustering assignments match the distribution of the data in the embedded space. This phase further refines the encoder weights to optimize both reconstruction and clusterability based on the combined loss function, expressed as:

\begin{equation} \label{mathref:combined-loss}
    \mathcal{L} = \alpha \mathcal{L}_{\text{recon}} + \beta \mathcal{L}_{\text{cluster}}
\end{equation}
where \( \alpha \) and \( \beta \) are hyperparameters that control the weight of the reconstruction and clustering losses, respectively.

The clustering loss for \acrshort{aeml} consists of multiple steps and is detailed in \cite{zhou2023deep}. Here we introduce its general form, defined as:
\begin{equation}
    \mathcal{L}_{\text{cluster}}^{\text{AEML}} = \text{CustomLoss}(Z)
\end{equation}
where \( Z \) is the set of embedded representations of the data points and \text{CustomLoss} is a function that measures clustering quality based on the relative distances between the samples.

The clustering loss for \acrshort{dcec}, which utilizes the Kullback-Leibler divergence, is detailed in \cite{guo2017deep}. However, we also briefly introduce it here for completeness. It is calculated as:
\begin{equation}
\mathcal{L}_{\text{cluster}}^{\text{DCEC}} = \text{KL}(P \| Q) = \sum_{i=1}^{n} \sum_{k=1}^{K} p_{ik} \log \frac{p_{ik}}{q_{ik}}
\end{equation}
where \( P \) is the target distribution, \( Q \) is the predicted distribution of the clusters, and \( p_{ik} \) and \( q_{ik} \) are the probabilities that point \( x_i \) belongs to cluster \( k \).

\section{EVALUATION METHODOLOGY} \label{sec:methodology}
In this section, we elaborate on the methodological details of the experiments carried out to assess the performance of the proposed architecture. First, we elaborate on the training data for model development, followed by a summary of the evaluation metrics used, and we end with considerations of the evaluation approach.

\subsection{TRAINING DATA} \label{sec:data}

To evaluate the performance of the proposed architecture we selected three real-world datasets, collected from research testbeds, in particular (i) the \acrfull{tcd} \cite{fontaine2019towards} containing DVB-T, LTE and WiFi transmissions, (ii) the \acrshort{its_l} dataset \cite{girmay2023technology} containing signals from 5 different \acrshort{rat}, i.e. LTE, 5G, WiFi, \acrshort{its}-G5 and C-V2X, and (iii) the LOG-a-TEC dataset \cite{gale2020automatic} comprised of LoRa, IEEE 802.15.4, SIGFOX and some proprietary technologies. This selection of three very diverse real-world datasets from the 868~MHz, 2.4~GHz and 5.9~GHz frequency bands enables the most extensive evaluation of unsupervised methods to date. 

\begin{figure}[ht]
    \centering
    \includegraphics[width=\linewidth]{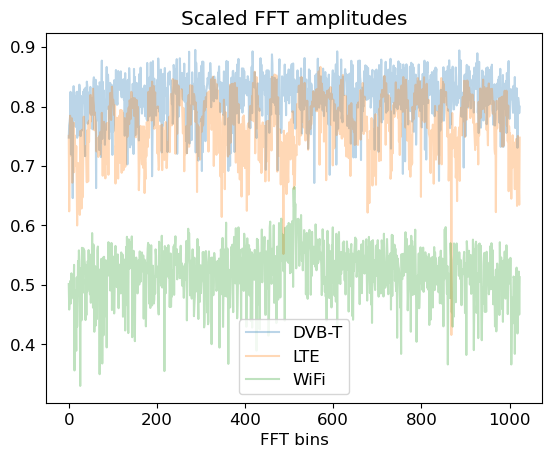}
    \caption{Samples for each signal type in the \acrshort{tcd} dataset.}
    \label{fig:samples}
 \end{figure}

\subsubsection{TECHNOLOGY CLASSIFICATION LABELED DATASET}

We purposely selected the labeled \acrfull{tcd} \cite{fontaine2019towards}, collected from different neighborhoods located in Ghent, Belgium. During the data collection process, the transmission times and settings were also recorded, therefore this is a real-world labelled dataset. However, we only use the labels for evaluation and not for training. The data captures the influence of the different environments on the different wireless technologies that are operating in the  2.4~GHz band, namely LTE, WiFi, and DVB-T. 
In the original dataset, samples are collected in different bands specific to each technology, while here, we normalize them in common 1024 \acrshort{fft} bins as if they were coexisting in the same channel. Samples of the captured transmissions by each \acrshort{ssdc} are depicted in Figure \ref{fig:samples}, showing that each of the \acrshort{rat}s has its own characteristic shape in the \acrshort{fft} domain.

\subsubsection{INTELLIGENT TRANSPORTATION SYSTEMS LABELED DATASET}
\begin{figure}[ht]
    \centering
    \includegraphics[width=\columnwidth]{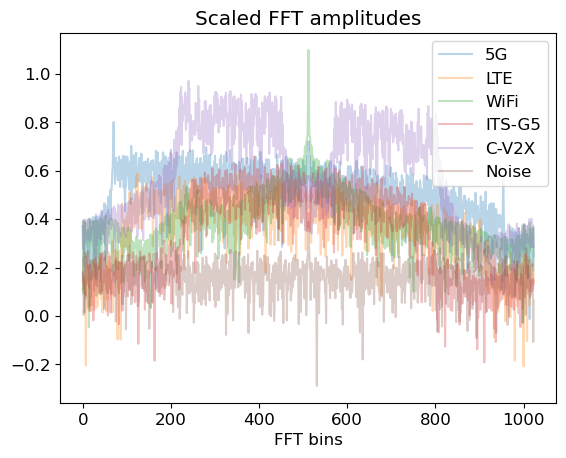}
    \caption{Samples for each signal type in the \acrshort{its} dataset (\acrshort{its_l}).}
    \label{fig:ITS_samples}
 \end{figure}

The \acrfull{its_l} dataset \cite{girmay2023technology} was collected in the region of Antwerp, Belgium. It contains signals from 5 different \acrshort{rat}s (LTE, 5G, WiFi, \acrshort{its}-G5 and C-V2X). The \acrshort{rat} technologies in the dataset are expected to coexist in the \acrfull{its} 5.9~GHz band. By including the \acrshort{its_l}, we expect to provide insights about the performance of the models when more types of \acrshort{rat} co-exist (i.e. 5 + noise compared to the 3 in the \acrshort{tcd}), from which some are significantly different regarding their shape (e.g. WiFi in blue and C-V2X in red in Figure \ref{fig:ITS_samples}), and some are very similar (e.g. LTE in green and 5G in orange in Figure \ref{fig:ITS_samples}).

\subsubsection{ULTRA NARROW BAND UNLABELED DATASET}
Finally, we also included one unlabeled, continuously sensed dataset collected from the LOG-a-TEC testbed \cite{gale2020automatic} in Ljubljana, Slovenia. This dataset consists of real-world spectrum traces in an ultra-narrow bandwidth of 192~kHz inside the European 868~MHz SRD band, which were sensed with a frequency of 5 \acrshort{psd} measurements per second, using 1024 \acrshort{fft} bins. According to an inspection of parts of the data, there are at least 4 technologies appearing in the recordings: LoRa, IEEE 802.15.4, SIGFOX, and some proprietary technology.  The goal of employing such data is to check the performance of the evaluated architectures in real-world scenarios, which exposes them to a significantly wider variety in the data that is being used, with all the artifacts and interference that occur during the acquisition. This makes the problem of \acrshort{ssdc} monitoring much more challenging and closer to the potential deployment environment.

\subsection{TRAINING CONFIGURATIONS} \label{sec:evaluation}
In this subsection, we provide the training configurations of the three architectures used in subsequent performance evaluation, the proposed \acrshort{ssdc} architecture and the two \acrshort{ae}-based architectures, \acrshort{aeml} and \acrshort{dcec}.

\subsubsection{CONFIGURATION OF THE CLUSTERING MODULE}
Evaluations are performed by measuring the K-means clustering performance on the embedded vectors provided by the models created with the proposed architecture \acrshort{ssdc} and two reference architectures, i.e., \acrshort{aeml} and \acrshort{dcec}, based on multiple metrics described in the following Section~\ref{sec:methodology}-\ref{sec:metrics}. We approached the experimental evaluation assuming no information about the exact types and quantity of existing \acrshort{rat}s in each of the datasets. Thus, for all three datasets, we set the upper number of possible \acrshort{rat}s to 10, which is higher compared to the actual number of classes $3$ and $6$ in the respective labeled datasets \acrshort{tcd} and \acrshort{its_l}, and approximately $5$ according to \cite{gale2020automatic} in the unlabeled dataset \acrshort{unb}. Selection of an approximate number of classes (i.e., operating \acrshort{rat}s) is also viable in real-world deployment based on the monitored frequency band, urban/rural area, etc. While such an approach closely imitates the real-world environment, following the problem formulation in Section~\ref{sec:problem}, it also sets fair ground for comparison of the evaluated models. In this way, we avoid potential bias induced by training and evaluation with the actual number of classes, considering that the \acrshort{ssdc} employs a clustering algorithm in its training pipeline, as detailed in Section~\ref{sec:architecture}-\ref{sec:ssdc_components}-\ref{sec:clus_module}, while the \acrshort{ae}-based models work without clustering feedback. Furthermore, a higher number of clusters may capture more details and variance within the data, potentially leading to more accurate representations of smaller, nuanced groups. However, this number (10 in our case) should not be too big since too many clusters might also mean overfitting to noise and outliers, making some clusters less meaningful.

\subsubsection{CONFIGURATION OF THE CNN MODULES}
To ensure fair comparisons across the three architectures, all models were constructed with the identical \acrshort{cnn} module, as proposed and detailed in Section~\ref{sec:architecture}-\ref{sec:ssdc_components}-\ref{sec:learning_module}. The implementation was written in Python programming language, using the Pytorch library.
The training of the \acrshort{ssdc} model was performed with the Adam optimizer with learning rate set to $10^{-3}$, and weight decay set to $10^{-5}$. The training iterations for the \acrshort{ssdc} were set to $250$ epochs. 
The \acrshort{ae}-based models \acrshort{aeml} and \acrshort{dcec} were configured following the description in Section~\ref{sec:architecture}-\ref{sec:ae_models} regarding the loss functions. The weight parameters $\alpha$ and $\beta$ were set to 1, which gives the same importance for both components of the combined loss function, expressed with Eq.~\eqref{mathref:combined-loss}.
The training was performed with the same configuration of the optimizer as for the \acrshort{ssdc} model. Regarding the training iterations, considering the \acrshort{ae}-based architectures work in two stages as described in Section~\ref{sec:architecture}-\ref{sec:ae_models}, the pre-training ran for $200$ epochs, and the second stage of joint-training ran for $50$ epochs, totaling $250$ epochs, same as for the \acrshort{ssdc} model. The embedded space dimensionality for all three models was set to $10$.

\subsection{CLUSTER QUALITY EVALUATION METRICS} \label{sec:metrics}

For better understanding and analyzing the embedded space, we provide three different types of visualizations of the clustering in the embedded space, namely \acrfull{tsne} \cite{van2008visualizing} of the embedded space, average of samples per cluster, and distribution of ground-truth classes per cluster.

Regarding the quantitative evaluation, we employ multiple existing metrics that are specific to the clustering approaches. According to the way the used metrics are calculated, they can be divided into two groups. The first group is based on labels, mainly evaluating the content/purity of the clusters regarding the different \acrshort{rat}s. The second group of metrics is based on the distances in the embedded space, mainly considering the inter-cluster and intra-cluster distances.

\subsubsection{VISUAL EVALUATION}

\acrshort{tsne} is useful for transforming high-dimensional data into a two-dimensional or three-dimensional space for visualization, thus revealing the structure and patterns in complex datasets. It is particularly suitable for examining the outcomes of the clustering algorithm and visualizing how various data classes are distributed. However, it is important to note that \acrshort{tsne} visualizations should be taken cautiously due to their nonlinear transformation and dimensionality reduction. They may not necessarily reflect the same data structure that exists in the higher-dimensional embedded space. Thus, the \acrshort{tsne} visualizations will be considered as informative and complementary views, supporting the cross-interpretation of multiple metrics when drawing conclusions for the performance of such self-supervised models.
Due to the large size of the data, the \acrshort{tsne} embedded space visualization is performed with a randomly sampled subset of $10,000$ points instead of the entire dataset. These visualizations aim to showcase the approximate structure of the clusters created by each model within the feature space.

Calculation and visualization of an average of the samples per cluster will show what the dominant type of \acrshort{rat} for each cluster is and enable the expert mapping of the identified clusters to actual \acrshort{rat}, as described in Section~\ref{sec:overview}-\ref{sec:phase4}. The averaging will suppress the less common types of \acrshort{rat}s and highlight the representative shape for each cluster. This visualization is particularly useful when unlabeled data is considered, which cannot be analyzed by the label-based metrics.

For the evaluation with labeled datasets, we also show the distribution of samples in each cluster regarding the ground-truth labels, providing one additional perspective of the clustering directly related to the label-based metrics described in the following section.

\subsubsection{LABELS-BASED METRICS}
Regarding the labels-based evaluation, we utilize 4 different metrics, namely \acrfull{nmi}, \acrfull{ari}, \acrfull{chs} and \acrfull{ccs}.

\acrlong{nmi} is a clustering metric that quantifies the mutual information between true class labels and cluster assignments while accounting for the scale of each. It ranges from 0 to 1, with higher values indicating better agreement between true classes and clusters. Given two cluster assignments \( U \) and \( V \), \acrshort{nmi} is defined as:
\begin{equation}
\text{NMI}(U, V) = \frac{2 \cdot I(U; V)}{H(U) + H(V)},
\end{equation}
where: \( I(U; V) \) is the mutual information between \( U \) and \( V \) and \( H(U) \) and \( H(V) \) are the entropies of \( U \) and \( V \), respectively.

\acrlong{ari} \cite{hubert1985comparing} metric compares the pairwise decisions (whether pairs of elements are in the same or different clusters) in the clustering outcome to the true labels, adjusting for chance grouping. Formally, it is expressed as:
\begin{equation}
\text{ARI} = \frac{\text{RI} - \mathbb{E}[\text{RI}]}{\max(\text{RI}) - \mathbb{E}[\text{RI}]},
\end{equation}
where \( \text{RI} \) is the Rand Index and \( \mathbb{E}[\text{RI}] \) is the expected value of the Rand Index.
It can range from -1 (indicating completely independent labeling) to 1 (perfectly matching labeling).

The \acrlong{chs} \cite{rosenberg2007v}, as a clustering metric, evaluates the quality of a clustering operation, assessing whether each cluster contains only members of a single class. It is defined as:
\begin{equation}
\text{Homogeneity} = 1 - \frac{H(C|K)}{H(C)},
\end{equation}
where \( H(C|K) \) is the conditional entropy of the classes given the cluster assignments, and \( H(C) \) is the entropy of the classes. The \acrshort{chs} has a range of values between $0$ and $1$. A score of $0$ is achieved when the clusters are completely mixed, meaning that each cluster contains an equal proportion of members from different classes.  A score of 1 is achieved when each cluster contains members from only one class and no class is spread across multiple clusters, representing perfect homogeneity.

The \acrlong{ccs} \cite{rosenberg2007v} is a metric evaluating the quality of clustering results in terms of how well it groups together elements of the same class. Completeness is computed based on the conditional entropy of the clusters given the class labels. It essentially measures whether all members of a given class are assigned to the same cluster, regardless of how many other classes are also present in that cluster, defined as:
\begin{equation}
\text{Completeness} = 1 - \frac{H(K|C)}{H(K)},
\end{equation}
where \( H(K|C) \) is the conditional entropy of the cluster assignments given the classes and \( H(K) \) is the entropy of the cluster assignments.  
\acrshort{ccs} score ranges from 0 to 1. A score of 0 indicates that the algorithm has dispersed members of a single class across multiple clusters, failing to group them together, and a score of 1 signifies that all members of each class are perfectly grouped within a single cluster.

All four labels-based metrics provide quantitative measures to evaluate the performance of clustering algorithms when true labels are known, which is the case for the \acrshort{tcd} and \acrshort{its_l} datasets. Each metric provides a different perspective of the clustering performance, considering the distribution of samples across the clusters relative to the ground-truth labels.

\subsubsection{DISTANCE-BASED METRICS} \label{sec:clusterquality}
For measuring the clustering performance based on the state in the embedded space, we combined metrics used in \cite{milosheski2023self} and \cite{cerar2023learning}, i.e., silhouette score, Davies-Bouldin score, and Calinski-Harabasz index.

The silhouette score \cite{rousseeuw1987silhouettes} is a metric used to measure the goodness of a clustering technique. It quantifies how well-separated clusters are in a dataset, ranging from $-1$ to $1$. A higher silhouette score indicates better-defined clusters, with scores closer to 1 indicating more cohesive and separated clusters, while scores near $0$ suggest overlapping clusters. A silhouette score that trends toward $-1$ suggests that many points in the dataset have been placed in inappropriate clusters.
Considering a sample \( i \), the average distance between \( i \) and all other points in the same cluster \( a(i) \), and the minimum average distance between \( i \) and points in a different cluster \( b(i) \), minimized over clusters, the silhouette score \( s(i) \) is given by:

\begin{equation}
s(i) = \frac{b(i) - a(i)}{\max(a(i), b(i))}.
\end{equation}

The overall silhouette score for the dataset is the mean silhouette score of all samples.

The Davies-Bouldin \cite{davies1979cluster} score is a clustering evaluation metric that measures the compactness and separation of clusters. Theoretically, this score ranges from $0$ upwards, with no fixed upper limit. Formal definition is as follows:
considering \( k \) clusters, with \( C_i \) being the centroid of cluster \( i \) and \( S_i \) being the average distance of all points in cluster \( i \) to the centroid \( C_i \), and similarity measure \( R_{ij} \) between clusters \( i \) and \( j \) defined as:

\begin{equation}
R_{ij} = \frac{S_i + S_j}{\|C_i - C_j\|},
\end{equation}
the Davies-Bouldin score \( DB \) is the average of the maximum \( R_{ij} \) for each cluster:

\begin{equation}
DB = \frac{1}{k} \sum_{i=1}^{k} \max_{j \neq i} R_{ij}.
\end{equation}

Low scores indicate good clustering quality, where clusters are compact (data points within clusters are close to each other) and well-separated (clusters are far apart from each other).

The Calinski-Harabasz \cite{calinski1974dendrite} index is a clustering evaluation metric that quantifies the ratio of variance between clusters to variance within clusters. The score can theoretically be as low as zero, indicating extremely poor clustering where the within-cluster variance is as high as the total variance. However, in practice, any non-trivial clustering will result in a score greater than zero. There is no theoretical upper limit to the score. Higher scores indicate better clustering performance, with more significant separation between clusters compared to the variance within clusters. Formal definition is as follows:
given \( k \) clusters and \( n \) data points, \( B_k \) as between-group dispersion matrix and \( W_k \) the within-group dispersion matrix, the Calinski-Harabasz index \( CH \) is defined as:

\begin{equation}
CH = \frac{\text{trace}(B_k) / (k - 1)}{\text{trace}(W_k) / (n - k)},
\end{equation}
where \(\text{trace}(B_k)\) is the trace of the between-group dispersion matrix and \(\text{trace}(W_k)\) is the trace of the within-group dispersion matrix.

The Davies-Bouldin, silhouette, and Calinski-Harabasz scores work based on the clustering result and inter-sample distances. Such metrics are crucial to provide a quality assessment of the clustering result when labels are absent, which is the case with the unlabeled dataset \acrshort{unb}. Furthermore, evaluation with Davies-Bouldin, silhouette, and Calinski-Harabasz can also provide valuable insights when working with labeled data because, in general, they quantify the quality of the clustering result based on the inter-cluster and intra-cluster distances, independent of the ground-truth labels.

\subsection{TRANSMISSION DETECTION EVALUATION METRIC} \label{sec:monitoring-evaluation}
Based on the mapping of the clusters to specific \acrshort{rat} classes, performed as described in Section~\ref{sec:overview}-\ref{sec:phase3}, we can evaluate the monitoring by calculating the standard multiclass classification evaluation metrics \textit{precision}, \textit{recall} and \textit{F1 score}. Since the problem of monitoring at this stage could be interpreted as multiclass classification, the evaluation with the aforementioned metrics is performed per class.

Thus, considering \( \mathcal{C} = \{1, 2, \ldots, K\} \) is the set of classes, for each class \( k \in \mathcal{C} \):
\begin{itemize}
    \item Precision \( P_k \) is the ratio of true positives to the sum of true positives and false positives for class \( k \):
    \begin{equation}
    P_k = \frac{\text{TP}_k}{\text{TP}_k + \text{FP}_k}.
    \end{equation}
    \item Recall \( R_k \) is the ratio of true positives to the sum of true positives and false negatives for class \( k \):
    \begin{equation}
    R_k = \frac{\text{TP}_k}{\text{TP}_k + \text{FN}_k}.
    \end{equation}
    \item F1 score for class \( k \) is the harmonic mean of precision and recall for class \( k \):
    \begin{equation}
    F1_k = \frac{2 \cdot P_k \cdot R_k}{P_k + R_k}.
    \end{equation}
\end{itemize}

This evaluation is performed using the models trained with the two labeled datasets, \acrshort{tcd} and \acrshort{its}, since these metrics necessitate ground-truth labels. For predicted labels, we use the identified labels for each formed cluster based on the visual evaluation. In the case of having clusters that contain multiple classes, we identify them based on the most dominant type of samples.

\subsection{COMPUTATIONAL PERFORMANCE METRICS}
As part of the evaluation, we also assess the potential deployment challenges of the models regarding their structure and size based on the number of trainable parameters and computational requirements based on the calculation of \acrfull{gflops}. \acrshort{gflops} is a measure of computational performance required or achieved by the model during the training and inference phases. It quantifies the number of floating-point operations (FLOPs) a model needs to execute per second in the billions (giga-). The calculations of the \acrshort{gflops} were carried out using the \textbf{fvcore}\footnote{\url{https://github.com/facebookresearch/fvcore}} library.

\begin{figure*}[ht!]
    \centering
    \begin{subfigure}[b]{0.29\linewidth}
        \centering
        \includegraphics[width=\linewidth]{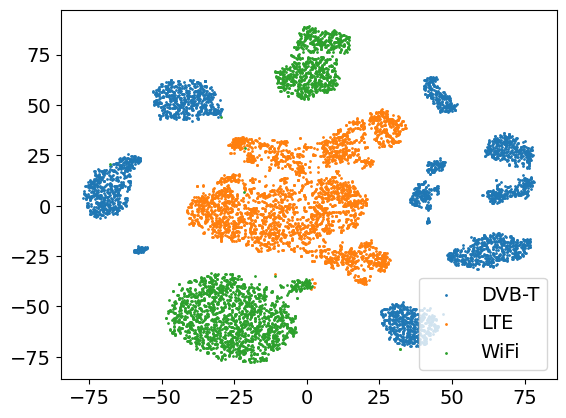}
        \caption{\acrshort{tsne} of ground-truth labels.}
        \label{fig:prop_fspaceGT}
    \end{subfigure}
    \begin{subfigure}[b]{0.29\linewidth}
        \centering
        \includegraphics[width=\linewidth]{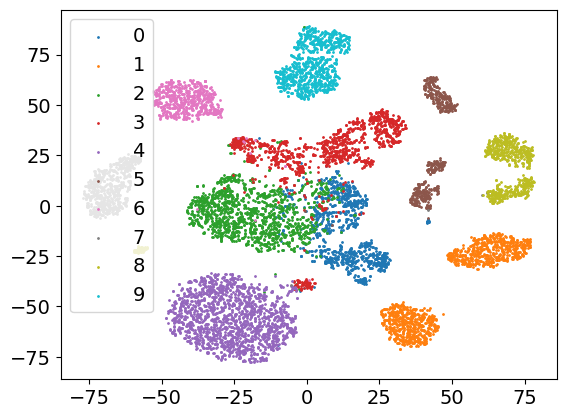}
        \caption{\acrshort{tsne} of learnt labels.}
        \label{fig:prop_fspaceKM}
    \end{subfigure}
    \begin{subfigure}[b]{0.39\linewidth}
        \centering
        \includegraphics[width=\linewidth]{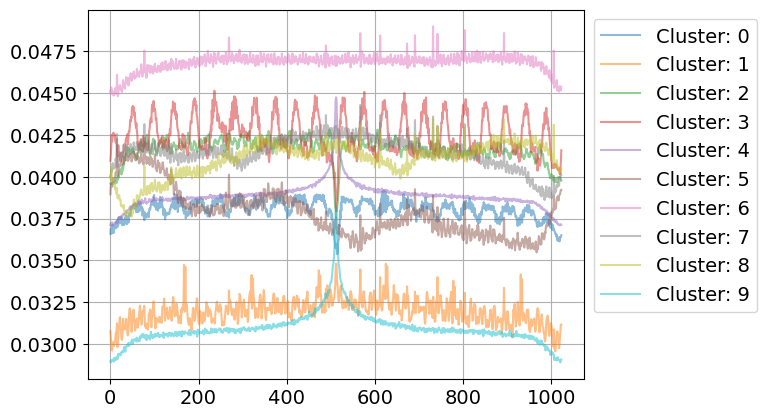}
        \caption{Average per cluster sample.}
        \label{fig:prop_avgC}
    \end{subfigure}
    \begin{subfigure}[b]{0.99\linewidth}
        \centering
        \includegraphics[width=\linewidth]{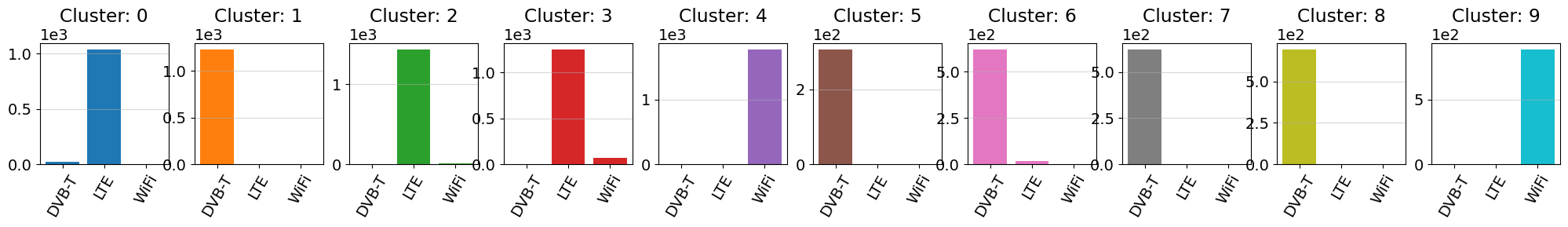}
        \caption{Per cluster distribution of ground-truth labels.}
        \label{fig:prop_clus_dist}
    \end{subfigure}
    \caption{Evaluation of the model learnt with the \acrshort{ssdc} architecture and \acrshort{tcd} dataset.}
    \label{fig:prop_eval_visual}
\end{figure*}
\begin{figure*}[ht!]
    \centering
    \begin{subfigure}[b]{0.29\linewidth}
        \centering
        \includegraphics[width=\linewidth]{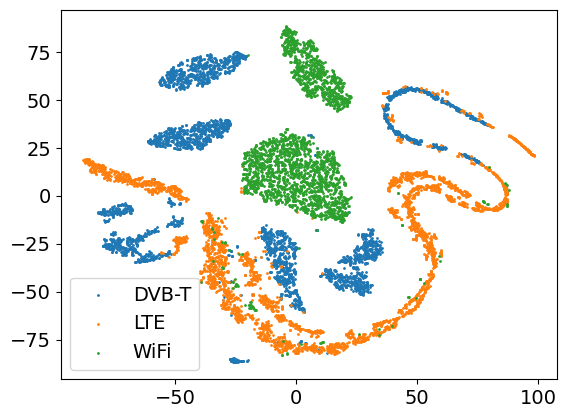}
        \caption{\acrshort{tsne} of ground-truth labels.}
        \label{fig:aeml_fspaceGT}
    \end{subfigure}
    \begin{subfigure}[b]{0.29\linewidth}
        \centering
        \includegraphics[width=\linewidth]{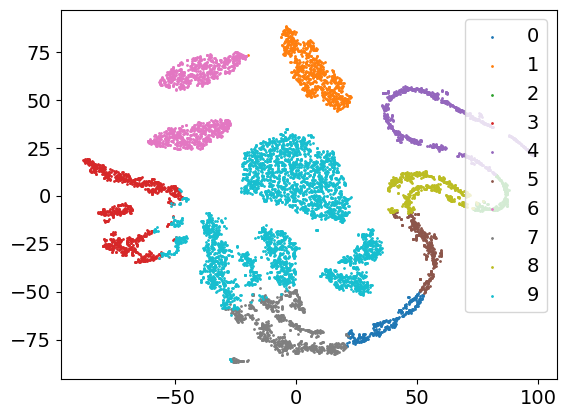}
        \caption{\acrshort{tsne} of the learnt labels.}
        \label{fig:aeml_fspace_KM}
    \end{subfigure}
    \begin{subfigure}[b]{0.39\linewidth}
        \centering
        \includegraphics[width=\linewidth]{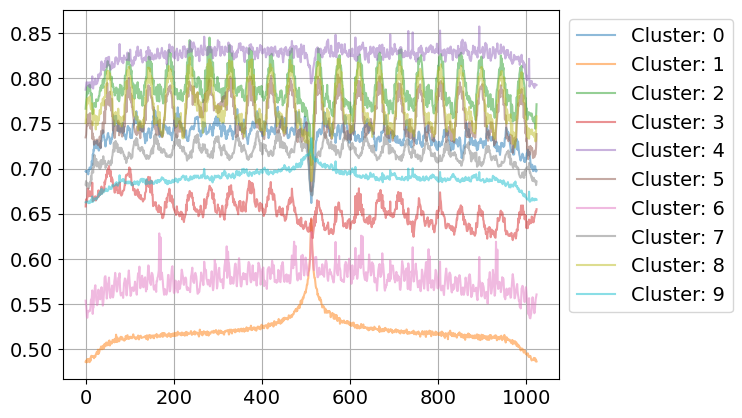}
        \caption{Average per cluster sample.}
        \label{fig:aeml_avg_clus}
    \end{subfigure}
    \begin{subfigure}[b]{0.99\linewidth}
        \centering
        \includegraphics[width=\linewidth]{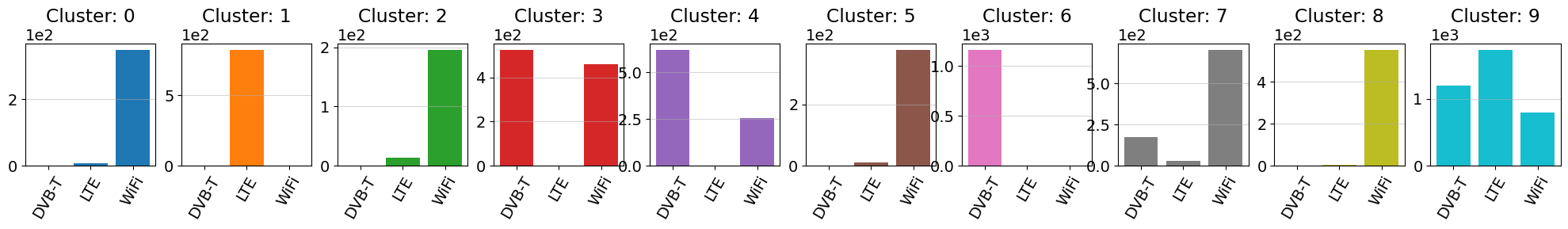}
        \caption{Per cluster distribution of ground-truth labels.}
        \label{fig:aeml_clus_dist}
    \end{subfigure}
    \caption{Evaluation of the model learnt with the \acrshort{aeml} architecture and \acrshort{tcd} dataset.}
    \label{fig:aeml_eval_visual}
\end{figure*}
\begin{figure*}[ht!]
    \centering
    \begin{subfigure}[b]{0.29\linewidth}
        \centering
        \includegraphics[width=\linewidth]{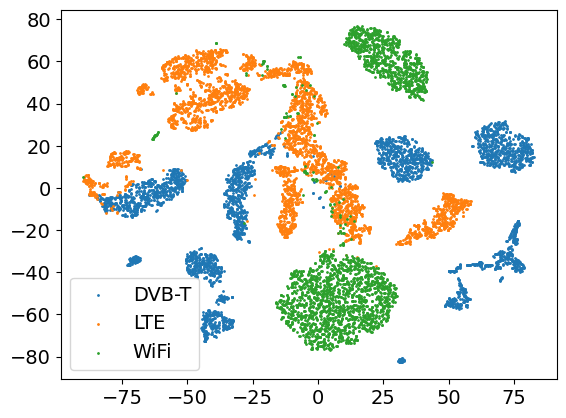}
        \caption{\acrshort{tsne} of ground-truth labels.}
        \label{fig:dcec_fspaceGT}
    \end{subfigure}
    \begin{subfigure}[b]{0.29\linewidth}
        \centering
        \includegraphics[width=\linewidth]{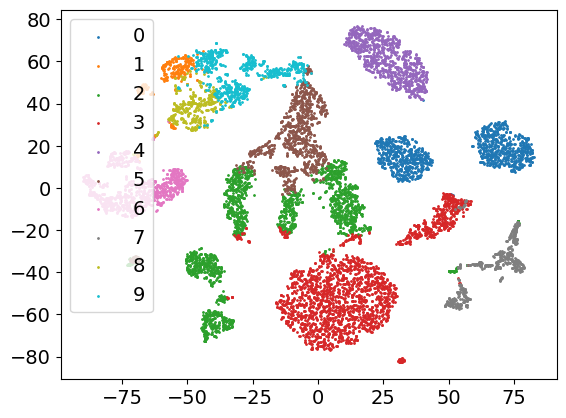}
        \caption{\acrshort{tsne} of the learnt labels.}
        \label{fig:dcec_fspaceKM}
    \end{subfigure}  
    \begin{subfigure}[b]{0.39\linewidth}
        \centering
        \includegraphics[width=\linewidth]{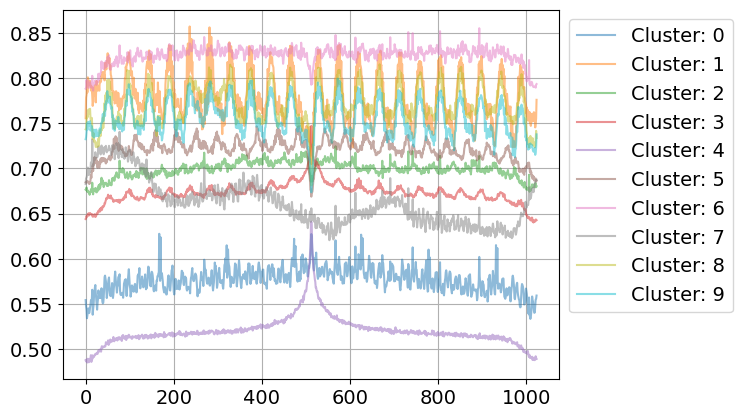}
        \caption{Average per cluster samples.}
        \label{fig:dcec_avg_clus}
    \end{subfigure}
    \begin{subfigure}[b]{0.99\linewidth}
        \centering
        \includegraphics[width=\linewidth]{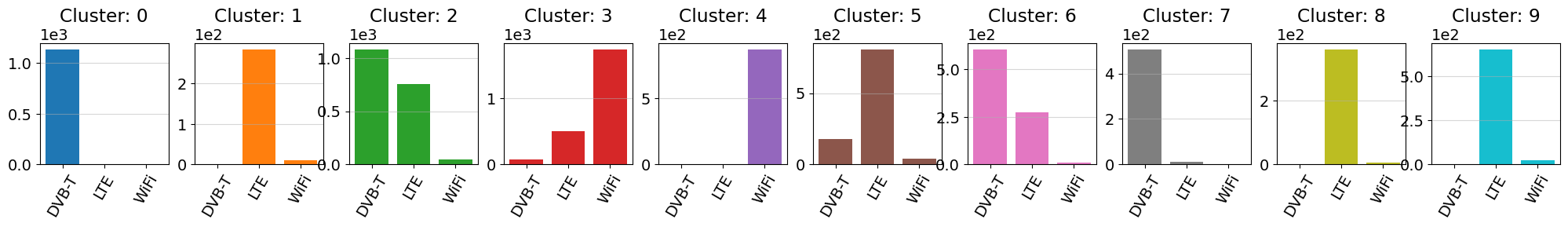}
        \caption{Per cluster distribution of ground-truth labels.}
        \label{fig:dcec_clus_dist}
    \end{subfigure}
    \caption{Evaluation of the model learnt with the \acrshort{dcec} architecture and \acrshort{tcd} dataset.}
    \label{fig:dcec_eval_visual}
\end{figure*}

\section{RESULTS} \label{sec:results}

The following results show the evaluation of models obtained from each of the three architectures, the proposed \acrshort{ssdc} and the \acrshort{aeml} and \acrshort{dcec} reference \acrshort{sota} architectures, with three different datasets, each of them addressing a specific challenge, as described in Section~\ref{sec:methodology}-\ref{sec:data}. 

\subsection{EVALUATION RESULTS WITH TCD-L} \label{sec:eval_tcd}

\subsubsection{VISUAL ANALYSIS}
Figures \ref{fig:prop_fspaceGT}, \ref{fig:aeml_fspaceGT} and \ref{fig:dcec_fspaceGT} depict the two-dimensional \acrshort{tsne} projected features learnt by each of the three models developed. The number of depicted clusters is 10, as selected in Section~\ref{sec:methodology}-\ref{sec:data}, while the colors represent the three types of \acrshort{rat}, namely DVB-T (blue), LTE (orange), and WiFi (green). From Figure \ref{fig:prop_fspaceGT}, it can be seen that while the model learns to create 10 clusters due to the clustering parameter selection, these clusters are well separated and homogeneous, with several of them containing the same technology. For instance, there are several blue clusters, all containing DVB-T, one large orange cluster containing LTE, and two large and one small green clusters containing WiFi. Although each model provides the same number of clusters, their shapes vary significantly. As can be seen in Figure \ref{fig:prop_fspaceGT}, the \acrshort{ssdc} model shows more tendency towards the creation of clusters with elliptic shapes, while the other two, the \acrshort{aeml} in Figure \ref{fig:aeml_fspaceGT} and \acrshort{dcec} in Figure \ref{fig:dcec_fspaceGT} have more variety in the cluster shapes. The distribution of the data input into the clustering methods, i.e., the distribution of the learnt embeddings in our case, is less suitable for well-shaped spheric clusters for the \acrshort{aeml} and \acrshort{dcec} models compared to the \acrshort{ssdc}. These two characteristics could be considered important advantages for the \acrshort{ssdc} model since it provides more "clustering-friendly" feature space.

In Figures \ref{fig:prop_fspaceKM}, \ref{fig:aeml_fspace_KM} and \ref{fig:dcec_fspaceKM}, we show the same feature space embedding, but this time colored with the 10 labels corresponding to the 10 clusters that were learnt by each model.
As can be seen in the figures, each of the formed clusters is largely homogeneous, consisting of samples mostly of a single \acrshort{rat} technology. Looking at the largest cluster in Figure \ref{fig:prop_fspaceKM}, we know from the ground-truth depiction in Figure \ref{fig:prop_fspaceGT} that it contains only LTE transmissions. It can be seen that it learnt to separate it into three different subclusters (0-blue, 2-green, and 3-red), where each of them contains samples from the same technology. This is also depicted by the clusters' distribution plot in Figure \ref{fig:prop_clus_dist}, with the same coloring as for the K-means result in Figure \ref{fig:prop_fspaceKM}. Thus, the averaged clusters' samples in Figure \ref{fig:prop_avgC}, also colored with the corresponding colors of the 10 clusters provided by the K-means, are aligned with this, showing the specific shapes for each of the technologies. The additional insight provided by the average plot is that the K-means groups the samples based on the \acrshort{rat} technology, where signal strength has a significant impact on the features, thus leading to the creation of separate clusters of the same technology.

Looking at the results of the reference autoencoder-based models depicted in Figures \ref{fig:aeml_eval_visual} and \ref{fig:dcec_eval_visual} for the \acrshort{aeml} and \acrshort{dcec} respectively, they seem to have more condensed clusters in the feature space compared to the proposed \acrshort{ssdc} model, which means that the samples of each technology in Figures \ref{fig:aeml_fspaceGT} and \ref{fig:dcec_fspaceGT} are less dispersed in the feature space. However, the fragmented and free forms of the clusters lead to greater confusion/misclassification when performing the K-means clustering. Thus, the resulting K-means clusters, according to Figures \ref{fig:aeml_fspace_KM} and \ref{fig:dcec_fspaceKM}, contain samples from different technologies. For example, cluster 3-red and cluster 9-cyan in Figure \ref{fig:aeml_eval_visual}, which are part of the elongated groups of samples, contain both DVB-T and LTE transmissions. A similar effect of clusters containing a mix of \acrshort{rat}s is even more prominent in the visualizations for the \acrshort{dcec} model instance, as can be seen in Figure \ref{fig:dcec_eval_visual}. Clusters learnt by the K-means algorithm in the \acrshort{dcec} provided feature space, visualized in Figure \ref{fig:dcec_fspaceKM} with assigned labels 2-green, 5-brown, and 6-pink, contain samples of DVB-T and LTE, and cluster 3-red contains LTE and WiFi, as can also be seen on the per cluster distribution of ground-truth labels in Figure \ref{fig:dcec_clus_dist}. This phenomenon is also leading to less distinguishable average samples in Figure \ref{fig:dcec_avg_clus}. For example, the shape of cluster 3-red, which has a dominant group of WiFi samples, does not have a smooth characteristic shape, which can be noticed for the average of cluster 4-purple, which has WiFi samples only.

\begin{table}[htbp]
\centering
\begin{threeparttable}
\centering
\caption{Evaluation of clustering by K-means on the different embedded spaces with \acrshort{tcd}.}
\label{tab:l_performance}
\begin{tabular}{|p{0.3\columnwidth}||m{0.15\columnwidth}|m{0.15\columnwidth}|m{0.15\columnwidth}|}
\hline
\cellcolor{gray!15}\diagbox[]{Metric}{Model} & \cellcolor{gray!15}\textbf{\acrshort{ssdc}} & \cellcolor{gray!15}\textbf{\acrshort{aeml}} & \cellcolor{gray!15}\textbf{\acrshort{dcec}} \\ \hline \hline
\cellcolor{blue!15}\acrshort{nmi} \color{Green} $\uparrow$ & \cellcolor{blue!15}\textbf{0.6209} & \cellcolor{blue!15}0.3266 & \cellcolor{blue!15}0.3998 \\ \hline
\cellcolor{blue!15}\acrshort{ari} \color{Green} $\uparrow$ & \cellcolor{blue!15}\textbf{0.3874} & \cellcolor{blue!15}0.1362 & \cellcolor{blue!15}0.23 \\ \hline
\cellcolor{blue!15}\acrshort{chs} \color{Green} $\uparrow$ & \cellcolor{blue!15}\textbf{0.9465} & \cellcolor{blue!15}0.4595 & \cellcolor{blue!15}0.5897 \\ \hline
\cellcolor{blue!15}\acrshort{ccs} \color{Green} $\uparrow$ & \cellcolor{blue!15}\textbf{0.4620} & \cellcolor{blue!15}0.2533 & \cellcolor{blue!15}0.3024 \\ \hline
\cellcolor{orange!15}Silhouette \color{Green} $\uparrow$ & \cellcolor{orange!15}0.3174 & \cellcolor{orange!15}\textbf{0.6050} & \cellcolor{orange!15}0.4325 \\ \hline
\cellcolor{orange!15}Davies-Bouldin \color{Green} $\downarrow$ & \cellcolor{orange!15}1.8727 & \cellcolor{orange!15}\textbf{0.528}6 & \cellcolor{orange!15}1.0396 \\ \hline
\cellcolor{orange!15}Calinski-Harabasz \color{Green} $\uparrow$ & \cellcolor{orange!15}18029 & \cellcolor{orange!15}\textbf{1259829} & \cellcolor{orange!15}357440 \\ \hline
\end{tabular}
\begin{tablenotes}
    \item Legend: \color{Green} $\uparrow$ \color{Black}- higher the better, \color{Green}{$\downarrow$} \color{Black}- lower the better. \\
    \colorbox{blue!15}{ } Labels-based metrics, \colorbox{orange!15}{ } Distance-based metrics
    
\end{tablenotes}
\end{threeparttable}
\end{table}

\begin{table}[htbp]
\centering
\begin{threeparttable}
\caption{Performance of different models on multiclass classification.}
\label{tab:monitor-eval-tcd}
\begin{tabular}{|p{0.15\columnwidth}|p{0.15\columnwidth}||m{0.10\columnwidth}|m{0.10\columnwidth}|m{0.10\columnwidth}|}
\hline
\cellcolor{gray!15}\textbf{Metric} & \cellcolor{gray!15}\textbf{Class} & \cellcolor{gray!15}\textbf{SSDC} & \cellcolor{gray!15}\textbf{AEML} & \cellcolor{gray!15}\textbf{DCEC} \\ \hline \hline
\multirow{3}{*}{\cellcolor{blue!15}} & \cellcolor{blue!15}DVB-T & \cellcolor{blue!15}\textbf{0.99} & \cellcolor{blue!15}0.76 & \cellcolor{blue!15}0.89 \\ \cline{2-5} \cellcolor{blue!15}Precision
& \cellcolor{blue!15}LTE & \cellcolor{blue!15}\textbf{0.97} & \cellcolor{blue!15}0.89 & \cellcolor{blue!15}0.81 \\ \cline{2-5} \cellcolor{blue!15}
& \cellcolor{blue!15}WiFi & \cellcolor{blue!15}\textbf{1.00} & \cellcolor{blue!15}0.57 & \cellcolor{blue!15}0.94 \\ \hline
\multirow{3}{*}{\cellcolor{blue!15}} & \cellcolor{blue!15}DVB-T & \cellcolor{blue!15}\textbf{1.00} & \cellcolor{blue!15}0.63 & \cellcolor{blue!15}0.79 \\ \cline{2-5} \cellcolor{blue!15}Recall
& \cellcolor{blue!15}LTE & \cellcolor{blue!15}\textbf{1.00} & \cellcolor{blue!15}0.59 & \cellcolor{blue!15}0.88 \\ \cline{2-5} \cellcolor{blue!15}
& \cellcolor{blue!15}WiFi & \cellcolor{blue!15}0.96 & \cellcolor{blue!15}\textbf{0.97} & \cellcolor{blue!15}0.96 \\ \hline
\multirow{3}{*}{\cellcolor{blue!15}F-score} & \cellcolor{blue!15}DVB-T & \cellcolor{blue!15}\textbf{1.00} & \cellcolor{blue!15}0.69 & \cellcolor{blue!15}0.84 \\ \cline{2-5} \cellcolor{blue!15}F-score
& \cellcolor{blue!15}LTE & \cellcolor{blue!15}\textbf{0.98} & \cellcolor{blue!15}0.71 & \cellcolor{blue!15}0.84 \\ \cline{2-5}\cellcolor{blue!15}
& \cellcolor{blue!15}WiFi & \cellcolor{blue!15}\textbf{0.98} & \cellcolor{blue!15}0.72 & \cellcolor{blue!15}0.95 \\ \hline
\end{tabular}
\begin{tablenotes}
    \item Higher is better.
\end{tablenotes}
\end{threeparttable}
\end{table}
\subsubsection{QUANTITATIVE ANALYSIS}

\paragraph{Evaluation with labels-based metrics}
As it was described in Section~\ref{sec:methodology}, besides the analysis based on visual evaluation of the clustering results, we also performed a quantitative evaluation based on multiple clustering-specific metrics. The results are summarized in Table \ref{tab:l_performance}, which lists the evaluation metrics in the first column, followed by the models obtained with the proposed architecture in column two and the reference autoencoder-based architectures in columns three and four, respectively. From the first row of Table \ref{tab:l_performance}, it can be seen that \acrshort{nmi} of the proposed \acrshort{ssdc} is almost double the value of the reference \acrshort{ae}-based models indicating stronger mutual dependence between the clustering results and the true labels, considering that 0 \acrshort{nmi} means no dependence. The clustering assignments produced by K-means on the \acrshort{ssdc}-provided feature-space more successfully identified the inherent groupings within the data that align closely with the ground truth categories, compared to the K-means assignments on the feature-space of the reference \acrshort{ae}-based models \acrshort{aeml} and \acrshort{dcec}. 

Considering the \acrshort{ari} metric values in Table \ref{tab:l_performance}, all three models have values above 0, meaning that their developed clusters are better than random clustering. Looking at each model's score separately, \acrshort{ssdc} leads by a margin of 0.14 to the second \acrshort{dcec}. This indicates that \acrshort{ssdc} has the greatest similarity between the clustering assignments and the ground-truth labels.
The \acrshort{chs} scores show very high values for \acrshort{ssdc}, meaning that the formed clusters contain mostly samples from a single class (\acrshort{rat}), while in the embedded spaces of the reference \acrshort{ae}-based models, there is much more variety in the clusters, supporting the conclusions made from the visual analysis.
Based on the \acrshort{ccs} metric, the proposed \acrshort{ssdc} is again much better (by a factor of 1.5 to 1.8). However, the values of all three models are in general low (the highest 0.4578 for  \acrshort{ssdc}) due to a large number of clusters ($10$) compared to the number of existing \acrshort{rat}s ($3$) in this dataset.

In general, \acrshort{ssdc} is superior according to the labels-based evaluation, \acrshort{dcec} is second, and the worst performance is achieved with the \acrshort{aeml} model.

\paragraph{Evaluation with distance-based metrics}
While all the labels-based metrics have shown a significant advantage for the proposed \acrshort{ssdc} model, the distance-based metrics silhouette, Davies-Bouldin and Calinski-Harabasz (last three rows in Table \ref{tab:l_performance}) show a clear advantage of \acrshort{aeml}. 
The higher silhouette score of the \acrshort{aeml} model indicates better-separated clusters, meaning that there is better cohesion of the inter-cluster samples and better separation between clusters, compared to \acrshort{ssdc} and \acrshort{dcec}. The advantage of better separation between the clusters and compactness of each cluster is already seen in Figures \ref{fig:prop_eval_visual}-\ref{fig:dcec_eval_visual}. The reference \acrshort{ae}-based models provide more dense clusters, with larger distances between them, while \acrshort{ssdc} has more dispersed clusters.
Davies-Bouldin and Calinski-Harabasz scores (last two rows in Table \ref{tab:l_performance}), which relate to the within-cluster dispersion and between-cluster separation, again indicate the more compact and better-separated clusters of the \acrshort{aeml} model compared to \acrshort{ssdc} and \acrshort{dcec}.

In general, both the labels-based and distance-based metrics are in line with the observations in the visual analysis of the embedded space (Figures \ref{fig:prop_eval_visual}, \ref{fig:aeml_eval_visual} and \ref{fig:dcec_eval_visual}) where better-separated feature space was provided by the \acrshort{aeml}. However, the clusters were much cleaner with the proposed \acrshort{ssdc} model mostly having single \acrshort{rat} per cluster. This means that the \acrshort{ssdc} model successfully learns more representative features for the different \acrshort{rat}s compared to the reference \acrshort{aeml} and \acrshort{dcec} models, and also exhibits better generalization capabilities, considering that the \acrshort{tcd} dataset is acquired across multiple environments.

\paragraph{Spectrum Monitoring Evaluation}
As described in Section~\ref{sec:methodology}-\ref{sec:monitoring-evaluation}, we evaluate the spectrum monitoring with the different models based on the identified classes, using the standard classification metrics, precision, accuracy, and F1-score. The results are summarized in Table \ref{tab:monitor-eval-tcd}. Precision, recall, and F1-score are calculated per class for the three existing classes in the \acrshort{tcd} dataset. The \acrshort{ssdc} model outperforms the \acrshort{ae}-based models by a significant margin, up to $40$ \acrshort{ppt}, in all but one case. There is a negligible difference in the Recall of the WiFi class, where the \acrshort{aeml} model has $1$ \acrshort{ppt} better performance over the other two, \acrshort{ssdc} and \acrshort{dcec}. This evaluation again confirms the superior generalization capabilities of the \acrshort{ssdc} model compared to the \acrshort{ae}-based models, considering that the \acrshort{tcd} dataset contains data from different environments, as described in Section~\ref{sec:methodology}-\ref{sec:data}, and the potential to use it for training a classifier without labeled data.

\subsection{EVALUATION RESULTS WITH ITS-L}

Following the methodology detailed in Section~\ref{sec:methodology}, in this paragraph, we analyze the results of the evaluation with the second labeled dataset \acrshort{its_l}. The evaluation is performed in the same order as for the previous \acrshort{tcd} dataset, with a focus on the quantitative analysis. 

\subsubsection{Evaluation with labels-based and distance-based metrics}
The performance of the models is depicted in Table \ref{tab:its_dataset_performance}, where again, the first four rows with blue background show the performance measured with the labels-based matrics, and the last three rows with orange background show the values of the three distance-based metrics. 

Regardless of the larger number ($6$) of different types of signals (5 \acrshort{rat}s + noise) in this dataset, the performance of all three models follows a similar pattern as in the evaluation with the \acrshort{tcd} dataset with $3$ \acrshort{rat}s. 
Regarding the labels-based metrics, the \acrshort{ssdc} model outperforms the other two across all four metrics.
\acrshort{dcec} is very close to the \acrshort{ssdc} on the labels-based metrics, whereas \acrshort{aeml} performs quite badly, which was not the case for the evaluation with the \acrshort{tcd} dataset. The relatively high performance of the \acrshort{dcec} means that the \acrshort{ae} structure is capable of capturing the different shapes of the signals. However, different loss functions used in the two \acrshort{ae}-based architectures have led to the creation of different structures in the embedded domain. The deep clustering component of the loss function of the \acrshort{aeml} model is more dominant compared to the reconstruction loss, leading to very compact clusters in the embedded domain while minimizing the influence of the shape of the samples in the learning process. This is confirmed by the high performance on the distance-based metrics of the \acrshort{aeml} model, providing compact and well-distanced clusters containing samples from different \acrshort{rat}s. The \acrshort{ssdc} model shows the most balanced performance, with comparably distinguishable clusters as the \acrshort{ae}-based models, which are also homogeneous, containing mostly single \acrshort{rat} samples. Overall, the \acrshort{ssdc} successfully captures the \acrshort{rat}-specific features when there is a larger number of \acrshort{rat}s, outperforming the \acrshort{ae}-based models.

\paragraph{Spectrum Monitoring Evaluation}
Spectrum monitoring performance of the three models on the \acrshort{its} dataset, regarding the precision, recall, and F1-score, is summarized in Table \ref{tab:model_performance}. The highest scores are bolded in the table. While the \acrshort{ssdc} model achieves the highest score in most cases across the evaluation with the three metrics across the six classes, it is evident that it has a very similar performance with the \acrshort{dcec} model. In general, we can notice a similar performance pattern as in the previous labels-based evaluation,  demonstrated in Table \ref{tab:its_dataset_performance}. This means that the \acrshort{ssdc} and \acrshort{dcec} can provide high-performance classifiers using completely unlabeled data, with only providing identification of the clusters, as part of the Phase 3, described in Section~\ref{sec:overview}-\ref{sec:phase3}.

\begin{table}[ht]
\centering
\begin{threeparttable}
\caption{Evaluation of clustering by K-means on different embedded spaces, \acrshort{its_l}.}
\label{tab:its_dataset_performance}
\begin{tabular}{|p{0.3\columnwidth}||m{0.15\columnwidth}|m{0.15\columnwidth}|m{0.15\columnwidth}|}
\hline
\cellcolor{gray!15}\diagbox[]{Metric}{Model} & \cellcolor{gray!15}\textbf{\acrshort{ssdc}} & \cellcolor{gray!15}\textbf{\acrshort{aeml}} & \cellcolor{gray!15}\textbf{\acrshort{dcec}} \\ \hline \hline
\cellcolor{blue!15}\acrshort{nmi} \color{Green}{$\uparrow$} & \cellcolor{blue!15}\textbf{0.8884} & \cellcolor{blue!15}0.3266 & \cellcolor{blue!15}0.8478 \\ \hline
\cellcolor{blue!15}\acrshort{ari} \color{Green}{$\uparrow$} & \cellcolor{blue!15}\textbf{0.7902} & \cellcolor{blue!15}0.1362 & \cellcolor{blue!15}0.7887 \\ \hline
\cellcolor{blue!15}\acrshort{chs} \color{Green}{$\uparrow$} & \cellcolor{blue!15}\textbf{0.9955} & \cellcolor{blue!15}0.4595 & \cellcolor{blue!15}0.9405 \\ \hline
\cellcolor{blue!15}\acrshort{ccs} \color{Green}{$\uparrow$} & \cellcolor{blue!15}\textbf{0.8022} & \cellcolor{blue!15}0.2533 & \cellcolor{blue!15}0.7718 \\ \hline
\cellcolor{orange!15}Silhouette \color{Green}{$\uparrow$} & \cellcolor{orange!15}0.5892 & \cellcolor{orange!15}0.6050 & \cellcolor{orange!15}\textbf{0.6819} \\ \hline
\cellcolor{orange!15}Davies-Bouldin \color{Green}{$\downarrow$} & \cellcolor{orange!15}0.6953 & \cellcolor{orange!15}\textbf{0.5286} & \cellcolor{orange!15}1.0396 \\ \hline
\cellcolor{orange!15}Calinski-Harabasz \color{Green}{$\uparrow$} & \cellcolor{orange!15}309780 & \cellcolor{orange!15}\textbf{1259829} & \cellcolor{orange!15}1102317 \\ \hline
\end{tabular}
\begin{tablenotes}
    \item Legend: \color{Green} $\uparrow$ \color{Black}- higher the better, \color{Green}{$\downarrow$} \color{Black}- lower the better. \\
    \colorbox{blue!15}{ } Labels-based metrics, \colorbox{orange!15}{ } Distance-based metrics
    
\end{tablenotes}
\end{threeparttable}
\end{table}

\begin{table}[htbp]
\centering
\begin{threeparttable}
\caption{Performance of different models on multiclass classification.}
\label{tab:model_performance}
\begin{tabular}{|p{0.15\columnwidth}|p{0.15\columnwidth}||m{0.10\columnwidth}|m{0.10\columnwidth}|m{0.10\columnwidth}|}
\hline
\cellcolor{gray!15}\textbf{Metric} & \cellcolor{gray!15}\textbf{Class} & \cellcolor{gray!15}\textbf{SSDC} & \cellcolor{gray!15}\textbf{AEML} & \cellcolor{gray!15}\textbf{DCEC} \\ 

\hline \hline \multirow{6}{*}{\cellcolor{blue!15}} & \cellcolor{blue!15}CV2X & \cellcolor{blue!15}0.94 & \cellcolor{blue!15}\textbf{1.00} & \cellcolor{blue!15}\textbf{1.00} \\ 

\cline{2-5} \cellcolor{blue!15}
& \cellcolor{blue!15}5G & \cellcolor{blue!15}\textbf{1.00} & \cellcolor{blue!15}0.99 & \cellcolor{blue!15}0.95 \\ 

\cline{2-5} \cellcolor{blue!15}Precision
& \cellcolor{blue!15}ITSG5 & \cellcolor{blue!15}\textbf{1.00} & \cellcolor{blue!15}0.89 & \cellcolor{blue!15}0.99 \\ 

\cline{2-5} \cellcolor{blue!15}
& \cellcolor{blue!15}LTE & \cellcolor{blue!15}\textbf{1.00} & \cellcolor{blue!15}0.93 & \cellcolor{blue!15}0.99 \\ 

\cline{2-5} \cellcolor{blue!15}
& \cellcolor{blue!15}WiFi & \cellcolor{blue!15}\textbf{1.00} & \cellcolor{blue!15}0.97 & \cellcolor{blue!15}0.98 \\ 

\cline{2-5} \cellcolor{blue!15}
& \cellcolor{blue!15}Noise & \cellcolor{blue!15}\textbf{1.00} & \cellcolor{blue!15}0.88 & \cellcolor{blue!15}\textbf{1.00} \\ 

\hline \multirow{6}{*}{\cellcolor{blue!15}Recall} & \cellcolor{blue!15}CV2X & \cellcolor{blue!15}\textbf{1.00} & \cellcolor{blue!15}0.99 & \cellcolor{blue!15}0.99 \\ 

\cline{2-5} \cellcolor{blue!15}
& \cellcolor{blue!15}5G & \cellcolor{blue!15}\textbf{1.00} & \cellcolor{blue!15}0.89 & \cellcolor{blue!15}0.98 \\ 

\cline{2-5} \cellcolor{blue!15}Recall
& \cellcolor{blue!15}ITSG5 & \cellcolor{blue!15}\textbf{0.96} & \cellcolor{blue!15}\textbf{0.96} & \cellcolor{blue!15}0.94 \\ 

\cline{2-5} \cellcolor{blue!15}
& \cellcolor{blue!15}LTE & \cellcolor{blue!15}0.99 & \cellcolor{blue!15}0.96 & \cellcolor{blue!15}\textbf{1.00} \\ 

\cline{2-5} \cellcolor{blue!15}
& \cellcolor{blue!15}WiFi & \cellcolor{blue!15}0.98 & \cellcolor{blue!15}0.83 & \cellcolor{blue!15}\textbf{0.99} \\ 

\cline{2-5} \cellcolor{blue!15}
& \cellcolor{blue!15}Noise & \cellcolor{blue!15}\textbf{1.00} & \cellcolor{blue!15}\textbf{1.00} & \cellcolor{blue!15}\textbf{1.00 }\\ 

\hline
\multirow{6}{*}{\cellcolor{blue!15}Fscore} & \cellcolor{blue!15}CV2X & \cellcolor{blue!15}0.97 & \cellcolor{blue!15}\textbf{0.99} & \cellcolor{blue!15}\textbf{0.99} \\ 

\cline{2-5} \cellcolor{blue!15}
& \cellcolor{blue!15}5G & \cellcolor{blue!15}\textbf{1.00} & \cellcolor{blue!15}0.94 & \cellcolor{blue!15}0.97 \\ 

\cline{2-5} \cellcolor{blue!15}F1-score
& \cellcolor{blue!15}ITSG5 & \cellcolor{blue!15}\textbf{0.98} & \cellcolor{blue!15}0.92 & \cellcolor{blue!15}0.96 \\ 

\cline{2-5} \cellcolor{blue!15}
& \cellcolor{blue!15}LTE & \cellcolor{blue!15}\textbf{1.00} & \cellcolor{blue!15}0.94 & \cellcolor{blue!15}0.99 \\ 

\cline{2-5} \cellcolor{blue!15}
& \cellcolor{blue!15}WiFi & \cellcolor{blue!15}\textbf{1.00} & \cellcolor{blue!15}0.90 & \cellcolor{blue!15}0.99 \\ 

\cline{2-5} \cellcolor{blue!15}
& \cellcolor{blue!15}Noise & \cellcolor{blue!15}\textbf{1.00} & \cellcolor{blue!15}0.94 & \cellcolor{blue!15}\textbf{1.00} \\ \hline
\end{tabular}
\begin{tablenotes}
    \item Higher is better.
\end{tablenotes}
\end{threeparttable}
\end{table}

\begin{figure*}[hbt!]
    \begin{subfigure}[b]{0.32\linewidth}
        \centering
        \includegraphics[width=\linewidth]{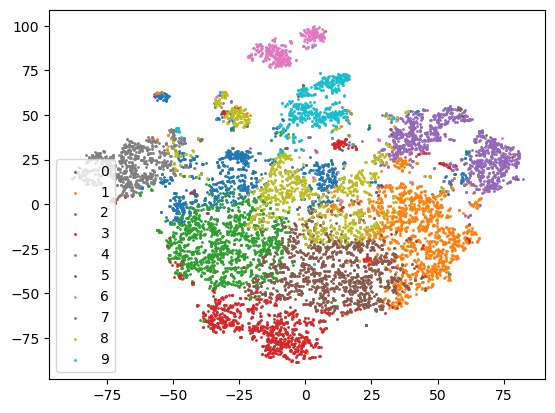}
        \caption{\acrshort{tsne} of learnt labels - \acrshort{ssdc}}
        \label{fig:prop_logatec_fspace}
    \end{subfigure}
    \begin{subfigure}[b]{0.33\linewidth}
        \centering
        \includegraphics[width=\linewidth]{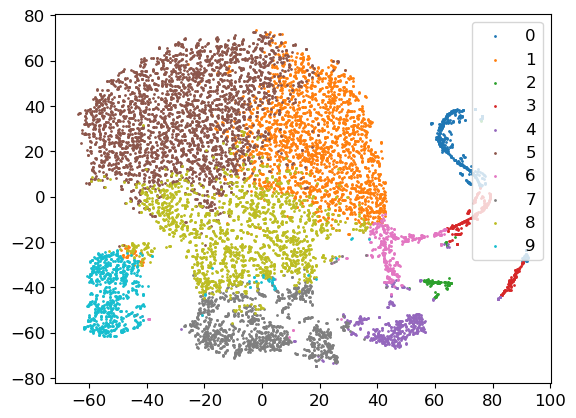}
        \caption{\acrshort{tsne} of the learnt labels -\acrshort{aeml}}
        \label{fig:aeml_logatec_fspace}
    \end{subfigure}
    \begin{subfigure}[b]{0.33\linewidth}
        \centering
        \includegraphics[width=\linewidth]{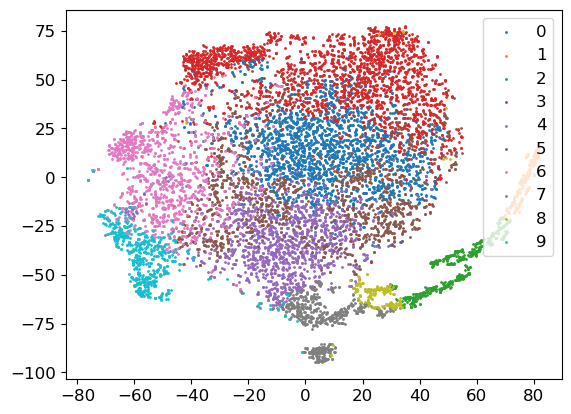}
        \caption{\acrshort{tsne} of the learnt labels -\acrshort{dcec}}
        \label{fig:dcec_logatec_fspace}
    \end{subfigure} \\
    \begin{subfigure}[b]{0.32\linewidth}
        \centering
        \includegraphics[width=\linewidth]{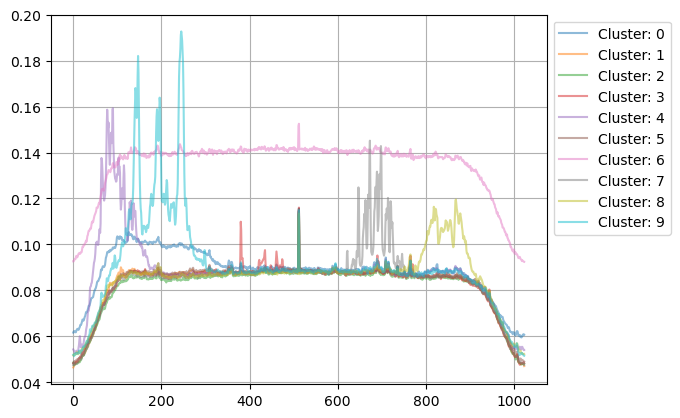}
        \caption{Average per cluster samples - \acrshort{ssdc}}
        \label{fig:prop_logatec_avg_clusters}
    \end{subfigure}
    \begin{subfigure}[b]{0.33\linewidth}
        \centering
        \includegraphics[width=\linewidth]{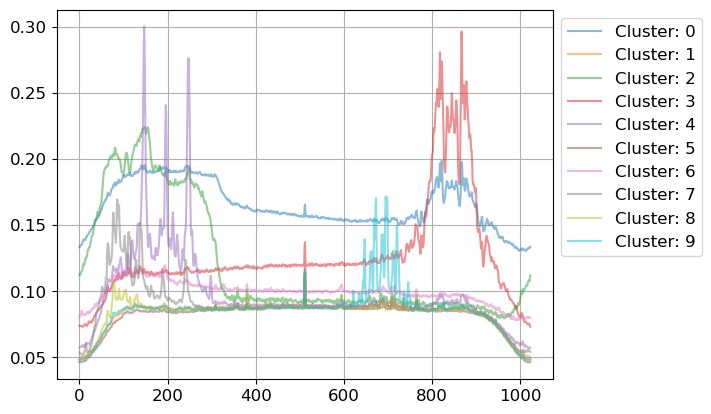}
        \caption{Average per cluster samples -\acrshort{aeml}}
        \label{fig:aeml_logatec_avg_clusters}
    \end{subfigure}
    \begin{subfigure}[b]{0.33\linewidth}
        \centering
        \includegraphics[width=\linewidth]{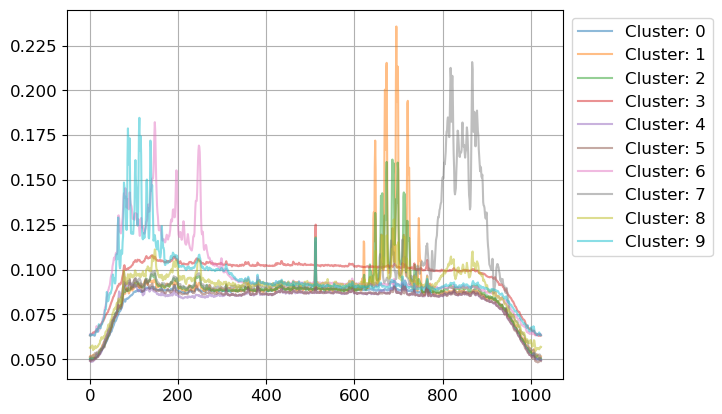}
        \caption{Average per cluster samples -\acrshort{dcec}}
        \label{fig:dcec_logatec_avg_clusters}
    \end{subfigure}
\caption{Feature-space and average clusters of the different models.}
\label{fig:data3_eval}
\end{figure*}

\subsection{EVALUATION RESULTS WITH UNB-U}

\subsubsection{Visual evaluation}
Figure \ref{fig:data3_eval} depicts the visual evaluation of the three models with the unlabeled dataset. In this case, since the ground-truth labels are not available, we can only analyze as part of the visual evaluation the  \acrshort{tsne} projections (Figures \ref{fig:prop_logatec_fspace}, \ref{fig:aeml_logatec_fspace} and \ref{fig:dcec_logatec_fspace}) and the cluster averages (Figures \ref{fig:prop_logatec_avg_clusters}, \ref{fig:aeml_logatec_avg_clusters} and \ref{fig:dcec_logatec_avg_clusters}). Considering that this dataset is acquired by continuous sensing, we expect it to have many noisy samples without clear distinction boundaries between them. Also, if there are significant amounts of samples with specific shapes, which would result from some specific \acrshort{rat} operating in the monitored band, we expect them to be grouped together in a separate cluster in the feature space.

Comparing the \acrshort{tsne} visualizations, the feature space of the \acrshort{ssdc} model appears to be separated into more distinctive and better-rounded clusters as shown in Figure \ref{fig:prop_logatec_fspace} compared to \acrshort{ae}-based models shown in Figures \ref{fig:aeml_logatec_fspace} and \ref{fig:dcec_logatec_fspace}. This makes the \acrshort{ssdc}-provided feature space advantageous for further analysis since it is easier to isolate each specific type of \acrshort{rat}, regardless of the type of clustering algorithm being used. Besides, the \acrshort{ae}-based models contain some well-separated clusters, such as the cluster 9-cyan in Figure \ref{fig:aeml_logatec_fspace}. There are also the elongated shapes that result in merged clusters that contain different \acrshort{rat}s, such as the clusters 3-red and 0-blue in Figure \ref{fig:aeml_logatec_fspace} and clusters 2-green and 8-yellow in Figure \ref{fig:dcec_logatec_fspace}. The same effect was also noticed in the evaluation with the \acrshort{tcd} labeled dataset in Section~\ref{sec:results}-\ref{sec:eval_tcd}.

These observations are also confirmed with the per-cluster averages depicted in Figures \ref{fig:prop_logatec_avg_clusters}, \ref{fig:aeml_logatec_avg_clusters} and \ref{fig:dcec_logatec_avg_clusters}. The well-distinguished clusters are easily recognizable shapes that are specific for different \acrshort{rat}s. This allows for further fine-grain analysis of the smaller clusters, which would not be possible for the other two, \acrshort{aeml} and \acrshort{dcec}, which show only 3-5 distinguishable clusters. Looking at the averages of the clusters of each model, all of them appear to have formed clusters for mostly similar transmission types. However, the merged clusters which were notable in the feature space visualization in Figures \ref{fig:aeml_logatec_fspace} and \ref{fig:dcec_logatec_fspace} are also influencing the per-cluster averages in Figures \ref{fig:aeml_logatec_avg_clusters} (e.g. cluster 0-blue and cluster 3-red) and \ref{fig:dcec_logatec_avg_clusters} (e.g. cluster 6-pink and cluster 9-cyan).

Based on analysis of the same dataset (\acrshort{unb}) in \cite{gale2020automatic}, we can tell which \acrshort{rat}s represent some of the formed clusters based on the shape of the per-cluster average. In Figure \ref{fig:prop_logatec_avg_clusters}, cluster 6-pink are IEEE 802.15.4 transmissions, clusters 7-gray and 9-cyan are two different proprietary transmissions, and cluster 4-purple are LoRA transmissions.

\subsubsection{Quantitative evaluation}
Quantitative evaluation of the three models with \acrshort{unb} dataset with the distance-based metrics is summarized in Table \ref{tab:ul_performance}. Results support the previous conclusions based on the visual analysis of the feature space and the per-cluster averages. The existence of some well-separated clusters has led the \acrshort{ssdc} model to significantly outperform the other two \acrshort{ae}-based reference models according to the silhouette score, i.e., \acrshort{aeml} by almost 40\% and \acrshort{dcec} by 150\%, which was not the case in the labeled datasets. It also shows comparable performance based on the Davies-Bouldin score by ranking second. The relatively good performance of \acrshort{ae}-based models results from the larger density of the formed clusters, which was also noticed in the evaluations with the labeled datasets. It is interesting to notice how the \acrshort{dcec} model performs significantly better on the Calinski-Harabasz score, which results from the good separation and density of some of the clusters on the margins of the feature space (i.e., cluster 1-orange, \ref{fig:dcec_logatec_fspace}).

Overall, judging by the distance-based metrics, all three models show comparable performance with the \acrshort{unb} unlabeled dataset. However, based on the visual evaluation, the clusters of the \acrshort{ssdc} model appear to form a larger number of recognizable transmission types; thus, the \acrshort{ssdc} model has more potential for further analysis because of the better segmentation of the feature space.

\begin{table}[ht]
\centering
\begin{threeparttable}
\centering
\caption{Evaluation of clustering by K-means on the different embedded spaces with \acrshort{unb}.}
\label{tab:ul_performance}
\begin{tabular}{|p{0.3\columnwidth}||m{0.15\columnwidth}|m{0.15\columnwidth}|m{0.15\columnwidth}|}
\hline
\cellcolor{gray!15}\diagbox[]{Metric}{Model} & \cellcolor{gray!15}\textbf{\acrshort{ssdc}} & \cellcolor{gray!15}\textbf{\acrshort{aeml}} & \cellcolor{gray!15}\textbf{\acrshort{dcec}} \\ \hline \hline
\cellcolor{orange!15}Silhouette \color{Green} $\uparrow$ & \cellcolor{orange!15}\textbf{0.2056} & \cellcolor{orange!15}0.1484 & \cellcolor{orange!15}0.0825 \\ \hline
\cellcolor{orange!15}Davies-Bouldin \color{Green} $\downarrow$ & \cellcolor{orange!15}1.5218 & \cellcolor{orange!15}\textbf{1.3183} & \cellcolor{orange!15}1.631 \\ \hline
\cellcolor{orange!15}Calinski-Harabasz \color{Green} $\uparrow$ & \cellcolor{orange!15}57037 & \cellcolor{orange!15}71520 & \cellcolor{orange!15}\textbf{363927} \\ \hline
\end{tabular}
\begin{tablenotes}
    \item Legend: \color{Green} $\uparrow$ \color{Black}- higher the better, \color{Green}{$\downarrow$} \color{Black}- lower the better..
\end{tablenotes}
\end{threeparttable}
\end{table}

\begin{table}[ht]
\centering
\caption{Computational complexity of the models.}
\label{tab:complexity_comparison}
\begin{tabular}{|p{0.25\columnwidth}||p{0.15\columnwidth}|p{0.25\columnwidth}|}
\hline
\cellcolor{gray!15}\diagbox[]{Metric}{Model} & \cellcolor{gray!15}\textbf{\acrshort{ssdc}} & \cellcolor{gray!15}\textbf{\acrshort{aeml}} / \textbf{\acrshort{dcec}} \\ \hline \hline
Num. params. & \textbf{128406} & 162827 \\ \hline
\acrshort{gflops} & \textbf{0.196672} & 0.38731776 \\ \hline
\end{tabular}
\end{table}

\subsection{POTENTIAL DEPLOYMENT CHALLENGES}
Regarding the architectural specifics, the \acrshort{ssdc} model could be considered easier to configure (number of training epochs and stopping criteria) compared to the reference \acrshort{ae}-based models because of the following main characteristics.

\subsubsection{Single loss function} Having two separate components that build the global loss function for the \acrshort{aeml} and \acrshort{dcec} could lead to conflicting situations in the learning process. For example, lowering the deep clustering part of the loss in the embedded domain (increasing the distance between samples) could lead to worsening the reconstruction loss. Additionally, in \acrshort{ae}-based models, there is no direct feedback from the actual classification process, such as the K-means for the \acrshort{ssdc} model, which means that actual classification performance could be even degraded by creating unnecessary clusters and fragmenting the feature space while improving the deep clustering and reconstruction loss.
  
\subsubsection{Inherently developed classifier} The training of the \acrshort{ssdc} inherently develops a classifier in parallel with the feature learning, which follows the assignments of the K-means algorithm. On the contrary, the \acrshort{ae}-based models require another classification algorithm to work on the extracted features since they only provide encoding of the input data without class information.

\subsubsection{Lower computational complexity}
As we described in Section~\ref{sec:methodology}, the reference \acrshort{ae}-based and the proposed \acrshort{ssdc} architectures utilize the same \acrshort{cnn} module. Consequently, during the inference mode (deployment scenario), all of the models will have a similar number of parameters. However, in the training mode, the \acrshort{ssdc} models have around 22\% fewer parameters than \acrshort{ae}-based models because of the encoder-decoder structure, which requires symmetrical \acrshort{cnn} module for decoding. Considering the \acrshort{gflops} calculation, as specified in Section~\ref{sec:methodology}, the proposed \acrshort{ssdc} architecture requires approximately half of the amount of \acrshort{gflops} compared to the \acrshort{ae}-based architecture. This means that tuning or retraining of the \acrshort{ssdc} model will require significantly less computational resources, as summarized in Table \ref{tab:complexity_comparison}. Smaller model sizes and lower computational requirements of the proposed \acrshort{ssdc} model could significantly benefit the potential deployment in more restricted devices closer to the edge of the network, thus extending the range of potential applications.

\section{CONCLUSION} \label{sec:conclusion}

In this paper, we investigated label-free RAT classification for spectrum sensing. We proposed a new spectrum sensing workflow, adequate for realistic set-ups that do not collect labels, based on a novel deep \acrfull{ssl}, that can be easily ported to various environments. This workflow enables the development of RAT classification models while avoiding the necessity of large labeled datasets. The SSL architecture is capable of autonomously learning low-dimensional features from raw FFT and enables the identification of different RATs without prior knowledge of the specific RATs in the monitored environment. The model is trained in a self-supervised manner, followed by manual cluster labeling by a human expert.

We evaluated the proposed architecture against state-of-the-art \acrfull{ae}-based architectures using three real-world datasets acquired from three different frequency bands, 2.4 GHz, 5.9 GHz, and  868 MHz, containing signals from over ten different \acrshort{rat}s. The trained model using the proposed architecture consistently outperformed the \acrshort{ae}-based models by up to 31 ppt F1 score according to spectrum monitoring evaluation, while at the same time requiring 22\% fewer trainable parameters and 50\% fewer \acrfull{flops}.

\section*{ACKNOWLEDGMENTS}
This work was funded in part from the Slovenian Research and Innovation Agency under the grant P2-0016 and in part from the European Union’s Horizon Europe Framework Programme under grant agreement No 101096456 (NANCY).
The NANCY project is supported by the Smart Networks and Services Joint Undertaking and its members.



\end{document}